\documentclass[final,5p,twocolumn,10pt,authoryear]{elsarticle}

\usepackage{amssymb}
\usepackage{amsmath}

\usepackage{enumitem}
\setitemize[1]{itemsep=0pt,partopsep=0pt,parsep=\parskip,topsep=0pt}
\setenumerate[1]{itemsep=0pt,partopsep=0pt,parsep=\parskip,topsep=0pt}

\usepackage{algorithmic}
\usepackage{multirow}
\usepackage{url}
\usepackage[dvipsnames]{xcolor}
\usepackage{amsfonts,amssymb}
\usepackage[caption=false,font=footnotesize,labelfont=sf,textfont=sf]{subfig}

\newtheorem{myDef}{Definition}

\usepackage[colorlinks]{hyperref}

\begin{document}

\begin{frontmatter}



\title{LWS: A Framework for Log-based Workload Simulation in Session-based SUT}



\author[tju]{Yongqi Han}

\author[tju]{Qingfeng Du\corref{cor1}}\ead{du\_cloud@tongji.edu.cn}

\author[tju]{Jincheng Xu}

\author[tju]{Shengjie Zhao}

\author[biz]{Zhekang Chen}

\author[biz]{Li Cao}

\author[tsu]{Kanglin Yin}

\author[tsu]{Dan Pei}

\address[tju]{{School of Software Engineering, Tongji University}, {201804}, {Shanghai}, {China}}

\address[biz]{{Bizseer}, {100083}, {Beijing}, {China}}

\address[tsu]{{Department of Computer Science and Technology, Tsinghua University}, {100084}, {Beijing}, {China}}

\cortext[cor1]{Corresponding author}

\begin{abstract}

Artificial intelligence for IT Operations (AIOps) plays a critical role in operating and managing cloud-native systems and microservice-based applications but is limited by the lack of high-quality datasets with diverse scenarios. Realistic workloads are the premise and basis of generating such AIOps datasets, with the session-based workload being one of the most typical examples. Due to privacy concerns, complexity, variety, and requirements for reasonable intervention, it is difficult to copy or generate such workloads directly, showing the importance of effective and intervenable workload simulation. In this paper, we formulate the task of workload simulation and propose a framework for Log-based Workload Simulation (LWS) in session-based systems. LWS extracts the workload specification including the user behavior abstraction based on agglomerative clustering as well as relational models and the intervenable workload intensity from session logs. Then LWS combines the user behavior abstraction with the workload intensity to generate simulated workloads. The experimental evaluation is performed on an open-source cloud-native application with both well-designed and public real-world workloads, showing that the simulated workload generated by LWS is effective and intervenable, which provides the foundation of generating high-quality AIOps datasets.

\end{abstract}




\end{frontmatter}


\section{Introduction}\label{sec:introduction}

Microservice-based applications composed by loosely coupled services have become the de-facto standard for delivering core business in large IT enterprises due to the independent and scalable nature~\citep{soldani2018pains}. With the development of cloud computing, microservice-based applications have the promise of migrating to the cloud-native architecture with scalability, resiliency, and elasticity~\citep{kratzke2017understanding}. Due to the heterogeneous, asynchronous, and independent nature, artificial intelligence for IT Operations (AIOps) that enhances the quality and reliability of IT service offerings~\citep{notaro2021survey} has been proposed and applied in the operation and management of cloud-native systems and microservice-based applications. Despite the steady growing efforts in various AIOps tasks such as anomaly detection and fault localization in recent years, AIOps datasets comprised of execution traces, monitoring metrics, logs, fault labels, and other related data commonly suffer from privacy restrictions, scale limitations, few data types and specific scenarios~\citep{li2022constructing}, which makes it difficult to guarantee the generality and real performance of AIOps efforts. Therefore, it is critical to continually generate high-quality AIOps datasets with diverse scenarios for the purpose of assisting and evaluating AIOps researches.

Generating high-quality AIOps datasets requires realistic workloads that represent incoming user requests in specific time intervals as the premise and basis. For commonly treated AIOps tasks such as anomaly detection and fault localization, realistic workloads can produce diverse failure scenarios better than simple constant workloads, which promotes the development of AIOps~\citep{li2022constructing}. One of the most typical workloads is the session-based workload describing the sequence of interdependent requests from the same user in session-based application systems~\citep{GosevaPopstojanova2006EmpiricalCO}. Generating such workloads in microservice-based applications is not a simple implementation of mocking user operations based on testing tools. Instead, it involves the following major challenges:

\begin{enumerate}[label={\textbf{\arabic*)}},leftmargin=0cm,itemindent=1cm,labelwidth=.5cm]

\item{\textbf{The Conflict between Realism and Privacy:} Realism in generating workloads is crucial for accurately reflecting real system performance in AIOps datasets. However, for privacy concerns, original real user operations are generally not directly accessible. Instead, observability data such as logs monitoring user operations needs to be converted to the workload specification for generation.}

\item{\textbf{Complexity and Variety:} Microservice-based applications benefit from the loose coupling characteristic, which enables them to provide more functionalities that can lead to complex user behavior combinations in workloads compared to monolithic applications. Besides, real workloads are usually highly volatile~\citep{reiss2012heterogeneity}, which vary considerably from constant or stepwise increasing workloads usually performed in sandboxed scenarios.}

\item{\textbf{Reasonable Intervention}: Some prior knowledge from other similar systems can be referred to in workload generation, such as holidays and events providing somewhat predictable shocks in a wide range of business scenarios~\citep{taylor2018forecasting}. The reasonable intervention based on prior knowledge in workload generation is beneficial for the analysis of predictable business patterns that happened in similar systems before.}

\end{enumerate}

\begin{figure}[!t]
\centering
\includegraphics[width=0.48\textwidth]{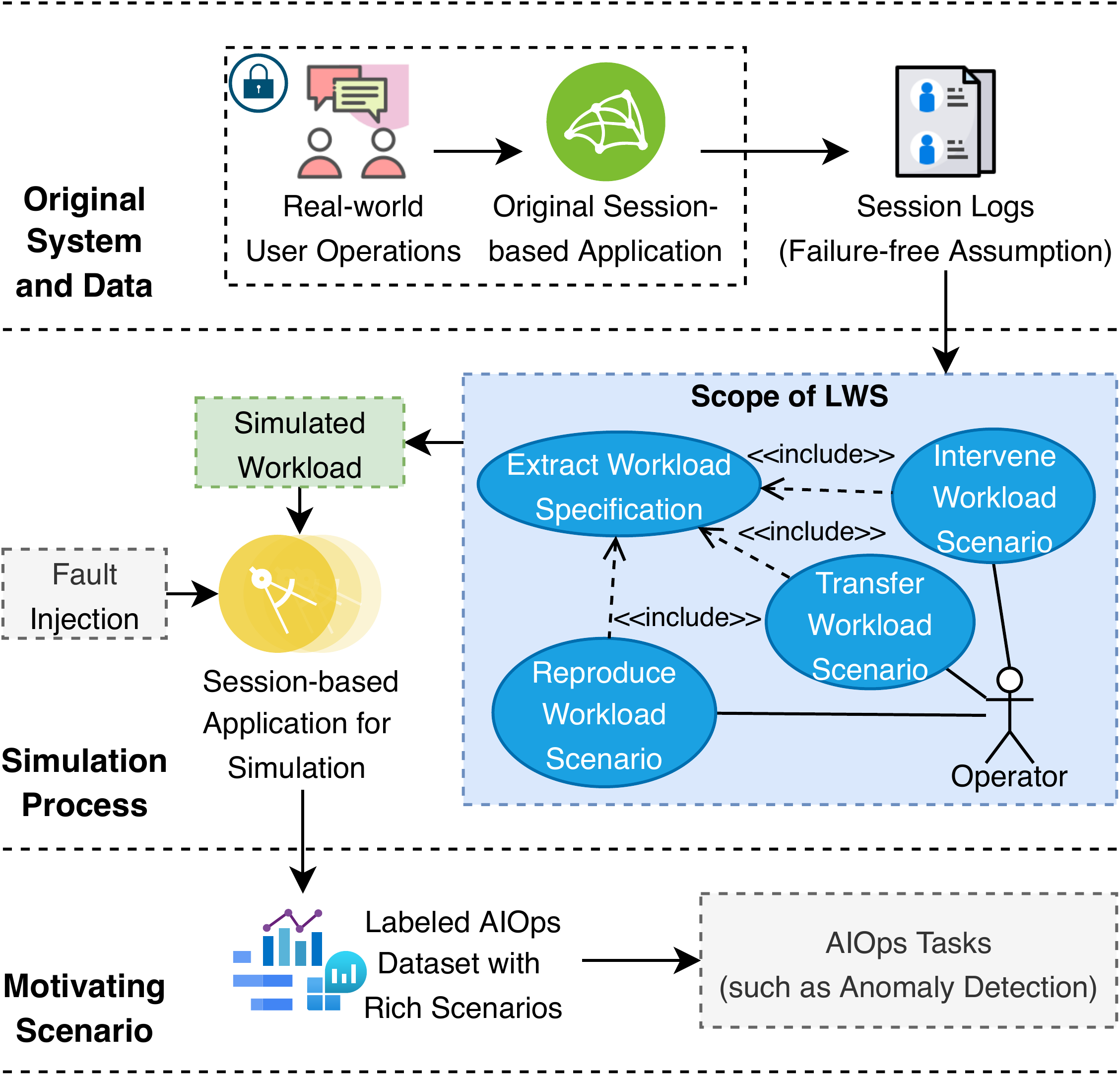}
\caption{The motivating scenario and typical use cases of LWS.}
\label{The motivating scenario and typical use cases of LWS}
\vspace{-4mm}
\end{figure}

These challenges reflect that the effective and intervenable simulation of workloads for generating AIOps datasets is more suitable and feasible than the simple copy of previous workloads based on original user operations. In response to these challenges, we propose an end-to-end framework for log-based workload simulation (LWS) to extract and generate workloads from logs produced by user operations that happened before in the session-based system under test (SUT). Figure \ref{The motivating scenario and typical use cases of LWS} shows the motivating scenario and typical use cases of LWS. The whole process of AIOps dataset generation and utilization is divided into three parts. First session logs from the original session-based application under real-world user operations are collected as the input of LWS. For potential privacy concerns, LWS is designed to require no additional data from the original system. In consideration of controllability and precise fault labels in AIOps datasets, we separate workloads from faults and assume that all failures in generated datasets are caused by fault injection. Therefore, user behaviors in such session logs are under the failure-free assumption and LWS only considers normal workloads. Then in the simulation process, the workload is simulated by LWS and executes on the session-based application for simulation along with injected faults. The scope of LWS is workload simulation containing typical use cases of reproducing, transferring and intervening workload scenarios as well as extracting workload specification as the basis, differing from the peak or average workload generation in performance testing. Finally, the labeled AIOps dataset with diverse scenarios is generated by collecting observability data in the session-based application for simulation and applied in AIOps tasks such as anomaly detection, which is the motivating scenario of LWS. To meet the above requirements, LWS extracts the workload specification including the user behavior abstraction based on agglomerative clustering as well as relational models and the intervenable workload intensity from session logs. Then LWS combines the user behavior abstraction with the workload intensity to generate simulated workloads. To summarize, the main contributions of our work are as follows:

\begin{itemize}

\item{We define the task of workload simulation formally with a set of necessary preliminaries. To the best of our knowledge, this is the first formal definition of the workload simulation task.}

\item{We propose an end-to-end framework named LWS for log-based workload simulation in session-based application systems. LWS extracts the workload specification including the user behavior abstraction and the intervenable workload intensity, combining them for simulated workload generation.}

\item{We perform a case study on an open-source cloud-native microservice demo application with both well-designed and real-world workloads to evaluate LWS comprehensively, demonstrating that LWS can generate effective and intervenable simulated workloads which provide the foundation of generating high-quality AIOps datasets..}

\end{itemize}

\section{Related Work}\label{sec:related work}

The key to workload simulation is the extraction of the workload specification. The workload specification describes key characteristics of user interactions with applications and is commonly summarized and explained by a workload model \citep{calzarossa2016workload}. We adopt the division in \cite{vogele2018wessbas} which divides the extraction of the workload specification into user behavior abstraction and workload intensity modeling. We group the related work into the above parts and explain the differences between LWS and existing approaches as follows.

\subsection{User Behavior Abstraction}

The user behavior sequence characterizes the single user session including which and how many operations a user finishes per session. One common approach is to configure a fixed sequence of user operations with manual scripts. This approach is simple but hardly varies characteristics and is far from real workloads \citep{Draheim2006RealisticLT}. For more representative user behavior abstraction, approaches based on analysis models are introduced, including approaches based on Markov Chains and Extended Finite State Machines (EFSMs). Markov-Chain-based models assume the memoryless property in the user behavior sequence, which are widely applied in web applications \citep{Li2003TestingTS}. Moreover, Customer Behavior Model Graphs(CBMGs) based on Markov chains which represent classes of users are extracted from logs in different systems \citep{Menasc1999AMF,Ruffo2004WALTyAU,parrott2020lodeston}. To overcome the limitation of modeling inter-request dependencies (also known as Guards and Actions, GaAs) in Markov-chain-based approaches, Shams et al. \citep{Shams2006AMA} propose an approach relying on EFSMs to describe valid user behavior sequences by predefined state variables as well as GaA parameters. To reduce the number of states in EFSMs, the kTails \citep{biermann1972synthesis} algorithm which merges pairs of equivalent states is also applied in the process of extraction \citep{goldstein2017experience}. WESSBAS \citep{vogele2018wessbas} combines approaches based on CBMGs and EFSMs, extracts GaAs from production session logs automatically, and generates executable workload models. Schulz et al. \citep{Schulz2019MicroserviceTailoredGO} and Barnert et al. \citep{barnert2021simulation} extend WESSBAS to microservice applications and databases by tailoring logs and modifying the Markov chain respectively. Besides, the stochastic form-oriented analysis \citep{Draheim2006RealisticLT,lutteroth2008modeling}, the probabilistic timed automata \citep{abbors2012mbpet}, the context-based sequential action \citep{Zhou2014LTFAM} and the latent features computed by URI space mapping vectors \citep{erradi2019web} are also applied in user behavior abstraction.

\subsection{Workload Intensity Modeling}

As introduced in \cite{curiel2018workload}, the workload intensity indicating the number of concurrent users can show daily or weekly access patterns and remarkable increases in short time spans. The constant or stepwise increasing workload intensity is widely used in performance testing frameworks such as Faban\footnote{http://faban.org/} and JMeter\footnote{https://jmeter.apache.org/} but cannot describe complex variations of the highly dynamic real-world workload intensity. To define and modify the workload intensity in a flexible and intuitive way, LIMBO \citep{v2014modeling} allows for modeling and configuring the workload intensity with load profiles and variables describing seasonal patterns, bursts, noise, and monotonic trends. Many approaches focus on workload intensity forecasting. \cite{herbst2014self,erradi2019web,fei2020elastic,feng2022fast} forecast the workload intensity by choosing suitable regression models including the Elastic Net regression model, the boosting decision tree regression prediction model, the Ridge regression model, the Lasso regression model, and the Multi-Layer Perceptron. Time series forecasting approaches including Telescope and Prophet are also applied to forecast the context-tailored workload intensity \citep{schulz2021context}. Besides, there also exist approaches replaying the workload directly \citep{fattah2020long}.

\subsection{Differences between LWS and Existing Approaches}

Most existing approaches only focus on workload specification extraction and the replay of original workloads. WESSBAS \citep{vogele2018wessbas} which extracts the relational model from HTTP request logs, generates the workload intensity based on LIMBO \citep{v2014modeling} and uses a domain-specific language (DSL) to specify workload models is a classic example and similar to our approach. Compared with these approaches, LWS has a fundamentally different goal of reproducing, transferring, and intervening workload scenarios for generating high-quality AIOps datasets with diverse scenarios rather than mere performance testing. Therefore, LWS focuses on quick, extensible, and intervenable workload generation instead of the peak or average workloads commonly used in WESSBAS and other approaches. As introduced in Section \ref{sec:lws framework}, LWS refers to the automatic GaA extraction process of WESSBAS in user behavior abstraction but abstracts the user behavior by a more general method and Hierarchical clustering. Moreover, LWS explores different strategies of workload intensity modeling and combines them for workload simulation which can be intervened effectively for quick, extensible, and intervenable workload generation. Besides, some other approaches \citep{herbst2014self,erradi2019web,fei2020elastic,schulz2021context,feng2022fast} attempt to forecast future workloads based on time series forecasting and workload specification. Different from these approaches, LWS provides an e2e framework for workload simulation rather than specific forecasts for concrete systems.

\section{Problem Definition}\label{sec:problem definition}

In this section, we present a set of formal descriptions for the task of workload simulation. First, we introduce the definition of user behaviors.

\begin{myDef}[\textbf{User Behavior}] 
Let $\mathcal{T}$ be a session-based SUT. A user behavior (denoted by $u$) is a single logical unit of work executed by the end user, which can access and possibly modify the contents in $\mathcal{T}$.
\end{myDef}

A session is a series of contiguous actions by a user on an individual system within a given time frame that is stored on the server side, but which action is a user behavior $u$ for workload simulation? To guarantee that $u$ can be properly selected, we propose the ASID properties. We do not claim these properties to be a complete set of desiderata, but they can be seen as a good starting point toward the understanding of $u$ in workload simulation.

\begin{enumerate}[label={\textbf{\arabic*)}},leftmargin=0cm,itemindent=1cm,labelwidth=.5cm]
   
 \item \textbf{Atomicity:} The entire $u$ takes place at once or does not happen at all. It will not occur partially. For example, an HTTP request can only be sent to $\mathcal{T}$ or not, and there is no midway. Hence, an HTTP request follows atomicity. On the contrary, a sequence of HTTP requests can be executed one by one in the beginning and not be executed anymore for the rest. Hence, a sequence of HTTP requests disobeys atomicity, and can not be seen as the user behavior. The atomicity property helps to determine the granularity of $u$.
 
  \item \textbf{Security:} The user behavior $u$ must not result in the failure of $\mathcal{T}$. For example, if a request leading to SQL injection attacks can crash the database, then $\mathcal{T}$ will not be able to process the subsequent requests normally. The security property makes sure that $\mathcal{T}$ can always process the user behaviors in the original and simulated workload. 
  
  \item \textbf{Isolation:} Multiple user behaviors can occur concurrently without leading to the inconsistency of the state of $\mathcal{T}$. Changes occurring in a particular $u$ will not be visible to any other $u$ until the particular change is written to memory or has been committed. Usually, the isolation property is provided by $\mathcal{T}$ for all the received requests. 
  
  \item \textbf{Durability:} Once a user behavior $u$ has completed execution, updates and modifications to $\mathcal{T}$ are stored and persist even if a system failure occurs. For example, the mouse-move event is usually meaningless for the system state and will not be stored by $\mathcal{T}$. Even though it is a necessary action executed by the user, it should not be seen as a user behavior. The durability property helps to exclude insignificant actions and focus on key user behaviors. 
  
\end{enumerate}

In a web-based application $\mathcal{T}$, a single HTTP request satisfying the ASID properties can be seen as a user behavior $u$. According to the request URL and the request method (GET/POST), user behaviors can be further divided into different types. We assume that parameters and their values do not affect the taxonomy of $u$ in this paper. Based on the definition of user behaviors, we define the user behavior sequence as follows:

\begin{myDef}[\textbf{User Behavior Sequence}]
Let $\mathcal{T}$ be a session-based SUT and $u_i$ be the $i^{\text{th}}$ user behavior within a session. A user behavior sequence denoted by $S$ is an ordered sequence of $(u_1, u_2, \cdots, u_l)$ which can be executed entirely without raising errors in~$\mathcal{T}$.
\end{myDef}

Consider the following case. Assume that $u_a$ is the user behavior of ``add\_to\_cart" and $u_b$ is ``placing\_order". The SUT $\mathcal{T}$ requires that $u_b$ cannot be executed alone and $u_a$ must always precede $u_b$. If we try to execute an ordered sequence of $(u_b, u_a)$ forcefully, then the sequence is definitely illegal and it may even lead to a system crash in $\mathcal{T}$. Hence, if a sequence could raise an error, we do not treat it as a user behavior sequence $S$.

In a web-based application $\mathcal{T}$ where a single HTTP request is $u$, $S$ can be seen as an ordered sequence of HTTP requests within a session. For simplicity, we refer to a user behavior sequence as a session, and we use both terms throughout the rest of the paper.

In workload simulation, what to do is important, and it has been guaranteed by $S$ as the action dimension. On the other hand, when to do is also important. The definition of workload intensity provides the time dimension:

\begin{myDef}[\textbf{Workload Intensity}]
\label{myDef-Workload-Intensity}
Let $I:t\rightarrow N^+$ be the function returning the number of active sessions at the timestamp of $t$. The workload intensity, denoted as $I$, is a sequence of discrete-time data indexed in time order whose y-value is $I(t)$.
\end{myDef}

In a real system, the number of active sessions is usually collected by the monitoring facilities with unceasing regularity, for example, every 15 seconds. Hence, we define x-value $t$ of the workload intensity $I$ on a discrete domain rather than a continuous domain.

Now we present the definition of workload simulation.

\begin{myDef}[\textbf{Workload Simulation}]
\label{myDef-Simulation}
Given a series of user behavior sequences $\Omega(S)=\{S_1, \cdots, S_k\}$ and the workload intensity $I$, $\Omega(S)$ will be executed on the session-based SUT $\mathcal{T}$ scheduled by $I$ to produce the original workload, and a set of logs will be collected. Workload simulation aims to produce a new workload containing characteristics of the original workload and reasonable interventions from existing real workload patterns.
\end{myDef}

It is worth mentioning that the task of workload simulation is not to achieve a 1:1 emulation of the original workload or just workload execution. On the one hand, a specific user behavior sequence $S$ is meaningless in a workload containing thousands of sessions. We are more interested in how a group of users interact with $\mathcal{T}$ rather than how a specific user interacts with $\mathcal{T}$. On the other hand, workload simulation is not the final purpose but a compulsory task for generating high-quality AIOps datasets with diverse scenarios. We may want to increase the number of users, improve the architecture of SUT or transfer the workload to a new environment. User behavior abstraction and workload intensity modeling will be more helpful.
 
More concretely, the simulated workload should be similar to the original workload in terms of user behaviors and the workload intensity while keeping $\mathcal{T}$ the same. Moreover, performance metrics collected by monitoring facilities should also keep similar between the original workload and the simulated workload. 

\section{LWS Framework}\label{sec:lws framework}

\begin{figure}[!t]
\centering
\includegraphics[width=0.5\textwidth]{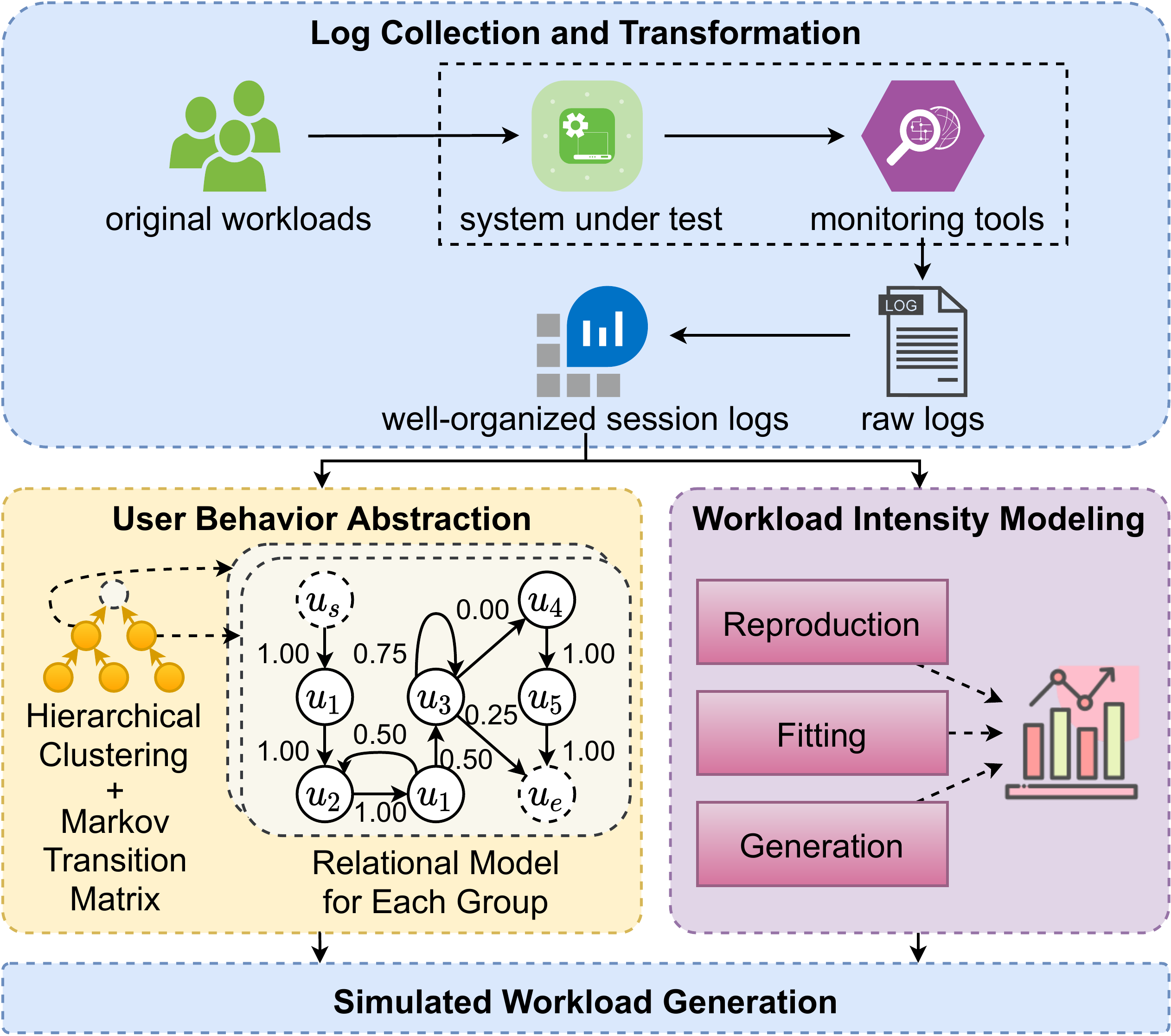}
\caption{The architecture of the LWS framework.}
\label{The architecture of the LWS framework}
\vspace{-4mm}
\end{figure}

In this section, we zoom in on the details of our proposed LWS framework. The architecture has been illustrated in Figure~\ref{The architecture of the LWS framework}, which consists of four different components:

\begin{itemize}
    \item \textbf{Log Collection and Transformation:} Collect and pre-process logs from the SUT under the original workload. 
    
    \item \textbf{User Behavior Abstraction:} Infer a high-level representation from thousands of sessions to describe user behaviors in a probabilistic way.
    
    \item \textbf{Workload Intensity Modeling:} Describe the number of active sessions indicating concurrent users as a time series for workload simulation.
    
    \item \textbf{Simulated Workload Generation:} Generate executable workloads based on user behavior abstraction and workload intensity. 
    
\end{itemize}

\subsection{Log Collection and Transformation}

A wide variety of monitoring facilities can be used for log collection. According to the deployment architecture of $\mathcal{T}$, developers or testers can take suitable solutions to monitor the original workload (as well as the simulated workload), especially in the cloud-native systems integrating with observability specifications such as OpenTelemetry\footnote{https://opentelemetry.io/} and tools such as Filebeat\footnote{https://github.com/elastic/beats} and Elasticsearch\footnote{https://github.com/elastic/elasticsearch} naturally.

A system can generate hundreds of logs in very little time, but not all the collected logs are necessary for workload simulation such as standalone logs recording event messages from the operating system. According to the definition of workload simulation, $\Omega(S)=\{S_0, \cdots, S_k\}$ and $I$ provide the action dimension and the time dimension as the necessary conditions. Therefore, the request-related logs to be filtered out are expected to contain the unique identifier to indicate a session (e.g., session ID), the request starting time, the request path, and the request parameters in a session-based application.

After the logs have been collected and filtered, we need to alter the structure and format of raw data to make it better organized for the analysis of the original workload. The useless fields will be removed to save storage space and reduce useless calculations. Under multiple concurrent requests, the collected logs are usually out of order, so we need to group them by the unique identifier and sort the logs by the request start time within each session. Now, the pre-processed logs are ready for the next steps.

\subsection{User Behavior Abstraction}\label{sec:user behavior abstraction}

\begin{figure}[!t]
\centering
\includegraphics[width=0.5\textwidth]{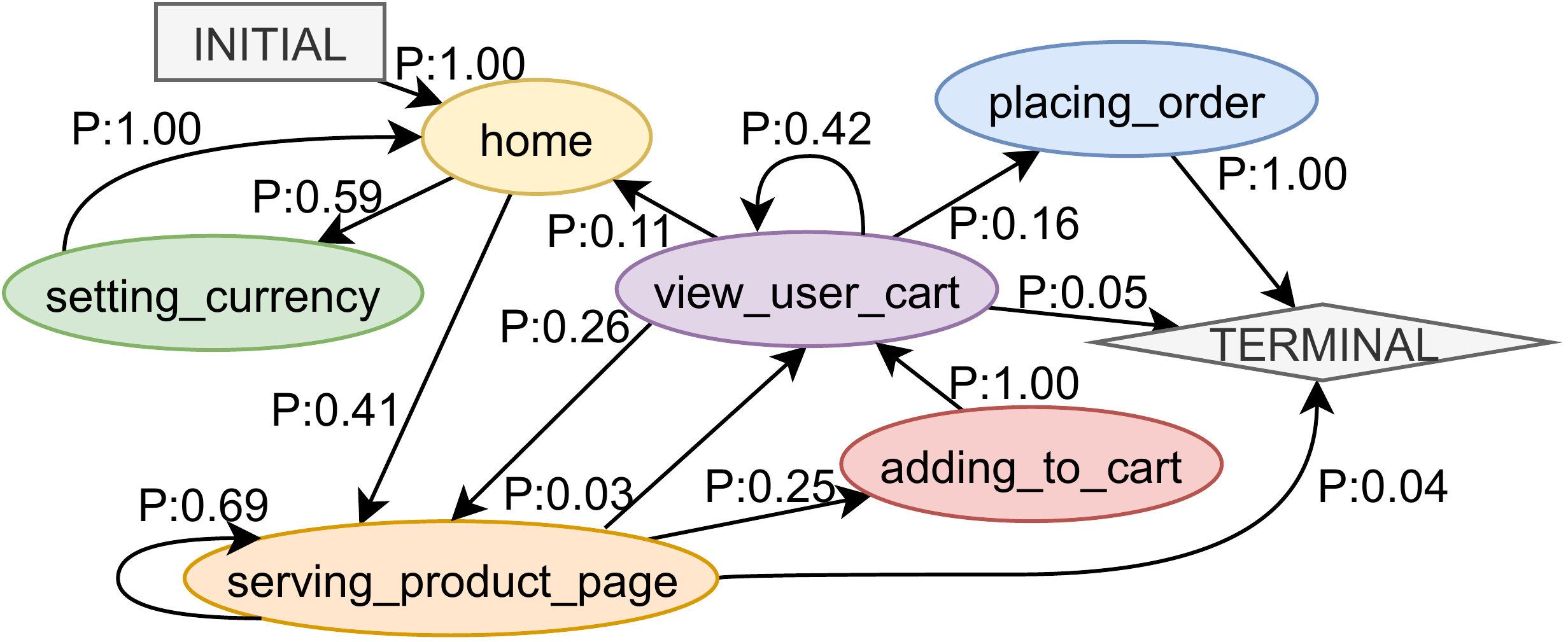}
\caption{The Markov probabilistic transition graph constructed on the dataset $\mathcal{A}$ in the case study.}
\label{Markov probabilistic transition graph}
\vspace{-4mm}
\end{figure}

Suppose there are $N_u$ types of user behaviors for $\mathcal{T}$. Inspired by Markov Chains, the simplest way to abstract user behaviors is to construct a $(N_u+2) \times (N_u+2)$ probabilistic transition matrix denoted by $\boldsymbol{A}$. The extra two columns/rows represent the dummy nodes of ``start" (denoted by $u_s$) and ``end" (denoted by $u_e$) for the start and end of a user behavior sequence, similar to the initial Markov state and the exit Markov state. Each row $\boldsymbol{A}_{r,\_}$ or each column $\boldsymbol{A}_{\_,c}$ corresponds to a certain type of user behavior, and each element $A_{r,c} \in \boldsymbol{A}$ is the transition probability from the user behavior $u_r$ to $u_c$. Since there will be no node preceding $u_s$, the transition probability from any node to $u_s$ must be 0. Similarly, since there will be no node following $u_e$, the transition probability from $u_e$ to any node must be 0. The extraction of $S$ starts from the node of $u_s$ and ends as long as the node of $u_e$ is selected. Refer to Figure~\ref{Markov probabilistic transition graph} for an example.

However, there are two main drawbacks in the above vanilla abstraction method as follows:

\begin{itemize}
    \item \textbf{The Impact of Different Groups of Users:} Real-world users typically have different preferences which can be divided into different groups. For example, some users just browse without purchasing, while others purchase their essential items. The vanilla abstraction method ignores differences in behaviors and proportions of different user groups.
    
    \item \textbf{The Neglect of Possible Temporal Invariants:} For example, an item has to be added to the cart (denoted by $u_{add\_to\_cart}$) before the customer places the order (denoted by $u_{placing\_order}$). However, in Figure~\ref{Markov probabilistic transition graph}, the behavior sequence $[u_s$, $u_{home}$, $u_{serving\_product\_page}$, $u_{view\_user\_cart}$, $u_{placing\_order}$, $u_e]$ is possible which violates the above restriction. The reason for this is that no temporal invariant has been taken into consideration when transiting between nodes.
    
\end{itemize}

To address these problems, we introduce a user behavior abstraction method, grouping users by Hierarchical clustering based on the Markov transition matrix and establishing a relational model for each user group containing pre-defined temporal invariants as follows.

\subsubsection{User Grouping}\label{User grouping}

To identify different groups of users, each user behavior sequence needs to be represented as feature embedding for (dis)similarity or ``distance'' computation. Inspired by the behavior model in WESSBAS, we calculate the $(N_u+2) \times (N_u+2)$ Markov probabilistic transition matrix $\dot{\boldsymbol{A}}$ for each user behavior sequence and flatten $\dot{\boldsymbol{A}}$ to a vector $\dot{\boldsymbol{v}}$ by Equation \ref{User behavior sequence embedding} as feature embedding where $\dot{A}_{r,c}$ is the element of $\dot{\boldsymbol{A}}$ in the $r^\text{th}$ row and the $c^\text{th}$ column and $\dot{v}_{i}$ is the $i^\text{th}$ element of $\dot{\boldsymbol{v}}$.

\begin{equation}\label{User behavior sequence embedding}
 \dot{A}_{r,c} = \dot{v}_{r\times (N_u+2)+c}
\end{equation}

Based on the above feature embedding, we can compute the user behavior sequence similarity and identify different user groups. Due to the natural hierarchy of user behaviors~\citep{kang2010detecting}, websites~\citep{lee2011novel} and user profiles~\citep{gu2020hierarchical}, we build on a typical hierarchical clustering algorithm, agglomerative clustering which adopts the bottom-up strategy to organize data into a tree structure~\citep{xu2005survey}. All $N_s$ user behavior sequences are divided into $N_s$ clusters at first. Then the cluster pair $(C_i, C_j)$ with the minimal variance calculated by Equation \ref{Agglomerative clustering} is found where $\dot{N}$ denotes the number of clusters, $C_i$ denotes the $i^\text{th}$ cluster, $\dot{\boldsymbol{v}}_h$ denotes the feature embedding of the $h^\text{th}$ user behavior sequence in the cluster, $\boldsymbol{m}_{C_p}$ denotes the central vector of the cluster $C_p$ and $\left \| \cdot \right \|$ denotes the two-norm of a vector.

\begin{equation}\label{Agglomerative clustering}
\begin{aligned}
\Delta (C_i, C_j) = \min_{1\leq p, q \leq \dot{N}} (\sum_{h \in C_p \cup C_q}\left \| \dot{\boldsymbol{v}}_h - \boldsymbol{m}_{C_p \cup C_q} \right \| ^2 - \\ \sum_{h \in C_p}\left \| \dot{\boldsymbol{v}}_h - \boldsymbol{m}_{C_p} \right \| ^2 - \sum_{h \in C_q}\left \| \dot{\boldsymbol{v}}_h - \boldsymbol{m}_{C_q} \right \| ^2)
\end{aligned}
\end{equation}

Then $C_i$ and $C_j$ are merged into a new cluster and the new cluster pair between the new cluster and the other clusters is searched for a new merger. The process is repeated until all user behavior sequences are in $N_c$ clusters where $N_c$ is the expected number of groups.

\subsubsection{Relational Model Establishing for Each User Group}

For each group of user behavior sequences identified in Section~\ref{User grouping}, we introduce Synoptic \citep{schneider2010synoptic} which generates a relational model to model user behaviors with temporal invariants. Three kinds of invariants are pre-defined and mined in user behavior sequences:

\begin{itemize}
    \item \textbf{$a$ Always Followed by $b$:} Whenever the behavior type $a$ appears, the behavior type $b$ always appears later in the same user behavior sequence. 
    
    \item \textbf{$a$ Never Followed by $b$:} Whenever the behavior type $a$ appears, the behavior type $b$ never appears later in the same user behavior sequence. 
    
    \item \textbf{$a$ Always Precedes $b$:} Whenever the behavior type $b$ appears, the behavior type $a$ always appears before $b$ in the same user behavior sequence. 
\end{itemize}

To mine the above invariants, each user behavior sequence is traversed once to collect the following counts:

\begin{itemize}
    \item $\forall a,$ OC[$a$] denoting the number of user behavior instances of $a$.
    
    \item $\forall a,b,$ FO[$a$][$b$] denoting the number of user behavior instances of $a$ which are followed by one or more user behavior instances of $b$.
    
    \item $\forall a,b,$ PR[$a$][$b$] denoting the number of user behavior instances of $n$ which are preceded by one or more user behavior instances of $a$.  
\end{itemize}

Then the invariants are mined by Equation ~\ref{Mine invariants from user behavior sequence}.

\begin{equation}\label{Mine invariants from user behavior sequence}
 \begin{split}
  a \textbf{ Always Followed by } b &\Leftrightarrow \text{FO}[a][b]=\text{OC}[a] \\
  a \textbf{ Never Followed by } b &\Leftrightarrow \text{FO}[a][b]=0 \\
  a \textbf{ Always Precedes } b &\Leftrightarrow \text{PR}[a][b]=\text{OC}[b]
 \end{split}
\end{equation}

\begin{figure}[!t]
 \footnotesize
 \begin{algorithmic}[1]
 \renewcommand{\algorithmicrequire}{\textbf{Input:}}
 \renewcommand{\algorithmicensure}{\textbf{Output:}}
 \REQUIRE initial graph $\mathcal{G}(\mathcal{V}, \mathcal{E})$; invariant set $\mathcal{I}$
 \ENSURE graph $\mathcal{G}(\mathcal{V}, \mathcal{E})$ satisfying all invariants in $\mathcal{I}$
 
 \textcolor{Green}{/* Step 1: Refinement */}
 \STATE set $\mathcal{CE}=findCounterexample(\mathcal{G}(\mathcal{V}, \mathcal{E}), \mathcal{I})$
 
 \WHILE{$\mathcal{CE}$ is not $\emptyset$}
  \STATE list $sp=[]$
  \FOR{$ce$ in $\mathcal{CE}$}
   \STATE $sp$.append$(findSplit(\mathcal{G}(\mathcal{V}, \mathcal{E}), ce))$
  \ENDFOR
  \IF{no valid split in $sp$}
   \STATE select a split $s$ arbitrarily
   \STATE $\mathcal{G}(\mathcal{V}, \mathcal{E}):=split(\mathcal{G}(\mathcal{V}, \mathcal{E}), s)$
  \ELSE
   \FOR{$s$ in $\textbf{sp}$}
    \STATE $\mathcal{G}(\mathcal{V}, \mathcal{E}):=split(\mathcal{G}(\mathcal{V}, \mathcal{E}), s)$
   \ENDFOR
  \ENDIF
  \STATE $\mathcal{CE}:=findCounterexample(\mathcal{G}(\mathcal{V}, \mathcal{E}), \mathcal{I})$
 \ENDWHILE
  
 \textcolor{Green}{/* Step 2: Coarsening */}
 \STATE bool $cm=\text{True}$
 \WHILE{$cm$}
  \FOR{$p,q$ in distinct vertex combinations of $\mathcal{V}$}
   \IF{$p\ne q$ and $p, q$ belong to the same type}
    \STATE $\mathcal{G}'(\mathcal{V}', \mathcal{E}')=merge(\mathcal{G}(\mathcal{V}, \mathcal{E}), p, q)$
    \STATE set $\mathcal{CE}'=findCounterexample(\mathcal{G}'(\mathcal{V}', \mathcal{E}'), \mathcal{I})$
    \IF{$\mathcal{CE}'$ is not $\emptyset$}
     \STATE $cm:=\text{False}$
    \ELSE
     \STATE $\mathcal{G}(\mathcal{V}, \mathcal{E}):=\mathcal{G}'(\mathcal{V}', \mathcal{E}')$
     \STATE $cm:=\text{True}$
     \STATE $\textbf{break}$
    \ENDIF
   \ENDIF
  \ENDFOR
 \ENDWHILE
 \RETURN $\mathcal{G}(\mathcal{V}, \mathcal{E})$
 \end{algorithmic}
 \caption{The refinement and coarsening algorithm}
 \label{Refinement and coarsening}
 \vspace{-4mm}
\end{figure}

After mining the invariants, the probabilistic transition matrix is also constructed as the adjacency matrix of the initial graph $\mathcal{G}(\mathcal{V}, \mathcal{E})$ where $\mathcal{V}$ denotes the vertex set and $\mathcal{E}$ denotes the edge set. Refinement and coarsening operations are performed in eliminating transition processes that violate the invariants (i.e. counterexamples) based on the initial graph as shown in Figure~\ref{Refinement and coarsening}. In the refinement process (lines 1-16), the Bisim algorithm \citep{schneider2010synoptic} is applied to eliminate counterexamples by generating new partition graphs. A partition is the splitting of the vertex in $\mathcal{V}$, belonging to the same user behavior type but corresponding to different instances. First $findCounterexample()$ attempts to find at least one counterexample for each invariant by traversing the vertices based on Breadth-First Search (BFS). For each counterexample, $findSplit()$ attempts to find a valid split that can eliminate the counterexample. The longest prefix of counterexample that does not violate any invariant is identified and the last vertex of the prefix is selected to be the candidate split. Then the new partition graph is generated based on the split. If there is no valid split but still exists counterexamples, the new partition graph is generated based on an arbitrary split. The above process is repeated until there exists no counterexample in the partition graph.

In the coarsening process (lines 17-32), the kTails algorithm \citep{biermann1972synthesis} is applied to reduce the number of partitions for the sake of brevity. Each pair of two different partitions belonging to the same user behavior type is chosen to merge. If there exist new counterexamples in the merged graph, the merge operation is rolled back. Finally, the most concise graph with no invariant violations is generated which is the abstraction of user behaviors.

Synoptic is not the only choice for abstraction. To meet the actual requirements of the SUT, more (or fewer) temporal invariants can be applied to improve the robustness of the generated relational model. For example, Perfume \citep{ohmann2014behavioral} proposes invariants with constrained resource measurements as well as WESSBAS proposes the invariant learning guards and actions (GaAs) which needs more business knowledge. However, it is impossible to find the exactly right and sufficient invariants because each invariant is designed for special business scenarios specific to the SUT. Expert knowledge of the SUT has a positive effect on the choice of invariants.

\subsection{Workload Intensity Modeling}

For better intensity modeling, we propose three methods to model the workload intensity $I$ from different perspectives, namely \textit{Reproduction}, \textit{Fitting} and \textit{Generation}.

\subsubsection{Reproduction}

In the reproduction method, we want to reproduce the time series of the original intensity $I$ exactly for workload simulation. In other words, the simulated intensity can be seen as the translation of the original intensity along the x-axis to a new time interval. The goal seems to be clear and straightforward, but it is almost impossible to achieve in reality because of the difference in session lengths (the number of user behaviors belonging to a session) or thinking times between the original workload and the simulated workload. For example, after an interval of 30 minutes, the original intensity may ramp down to 0 active sessions, but we cannot make sure that the sessions of simulated workload are also finished.

To simplify the problem, we follow the principle of Occam's Razor and only consider the start time of each session rather than the duration. That is, $I(t)$ denotes the number of sessions which starts in the time of $t$ rather than the number of active sessions in the reproduction method (as well as the other two methods). Suppose that the interval between the start time of the $i^\text{th}$ session and the $(i+1)^\text{th}$ session is $\triangle t$ in the original intensity. In the simulated intensity generated by the reproduction method, the interval between the start time of $S_i$ and $S_{i+1}$ is also expected to be $\triangle t$. 

To make sure that the domain of $I$, denoted by $t \in \lbrack t_{s}, t_{e} \rbrack$, can be divided into successive equally spaced points to be a standard time series, we set up an array of initially empty buckets. Each of them represents a time span of $\delta$ seconds, so from left to right each bucket covers the range of $[t_s, t_s+\delta), [t_s+\delta, t_s+2 \times \delta), \cdots, [t_e-\delta', t_e]$ where $\delta' = (t_e-t_s) \; mod \; \delta$ ($mod$ means the modulo operation). Then we go over the start time of each session and put each session in its bucket. In this way, $I(t)$ discards the fine-grained information on the start time of each session, but it can describe the overall intensity in the form of a standard time series.

If the original workload consists of $N_S$ sessions, we also have to extract $N_S$ user behavior sequences from the relational model to make sure that the total number of sessions is consistent. Consequently, we can reproduce the original intensity with respect to the session start time.

\subsubsection{Fitting}\label{Fitting}

\begin{figure}[!t]
\centering
\includegraphics[width=0.5\textwidth]{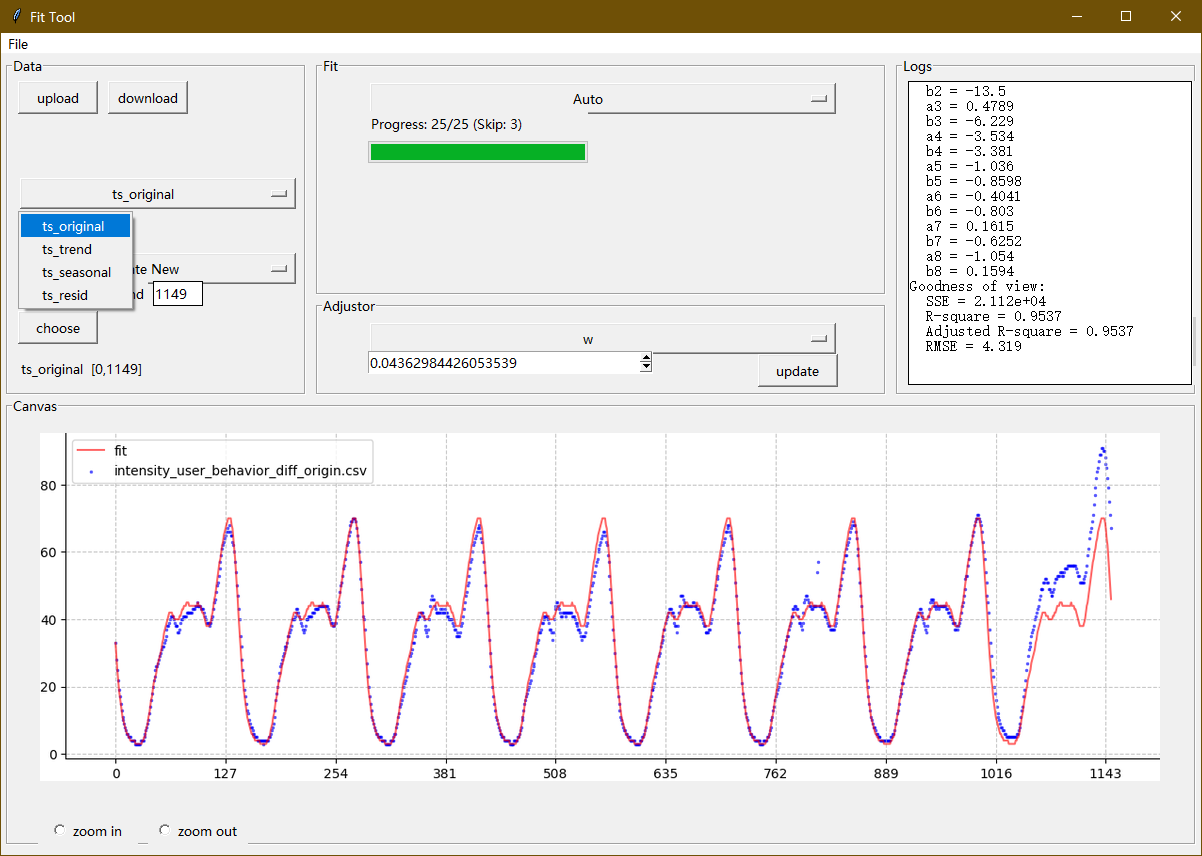}
\caption{The developed tool for the fitting method.}
\label{The developed tool for the fitting method}
\vspace{-4mm}
\end{figure}

Sometimes we are not only interested in reproducing the original intensity, but also interested in the performance of the SUT if we deliberately change the original intensity. For example, we may want to produce a workload that is longer than the original workload and increase the concurrent user capacity to see how the memory usage will change under the new workload. The new workload relies on the characteristics of the original workload rather than the simple translation. To achieve this goal, we propose the fitting method to learn the characteristics of the original intensity based on basic mathematical expressions. The first step is to decompose the original intensity into three distinct components including seasonality, trend, and noise based on time series decomposition algorithms such as RobustSTL~\citep{DBLP:conf/aaai/WenGS0XZ19}. This step is to make sure that we can take a deeper investigation into the characteristics of the original intensity. For example, the intensity in an online game may have a significant repeating short-term cycle on weekends, and we can analyze the seasonality independently to increase the peak on weekends but keep the online players on weekdays unchanged. The next step is to use a set of basic mathematical expressions such as polynomials to fit the targeted time series and choose the most suitable expression. Quasi-Newton methods~\citep{dennis1977quasi} are introduced to resolve the non-linear equations so that we can resolve the time series to a parameterized mathematical expression. At last, we can adjust the expression parameters to control the shape of the targeted time series freely.

For the convenience of user interaction when fitting the time series, we develop a tool with a visual interface of windows, as shown in Figure~\ref{The developed tool for the fitting method}. The intensity can be imported from a data file and subsequently decomposed into different components. It is worth mentioning that the decomposition algorithm usually requires the parameter of ``period" as the prior knowledge of the time series. To compute the period value, we perform the fast Fourier transformation and the largest peak in the frequency domain can indicate the possible period. Then, a decent interval rather than the whole range of the time series should be specified before fitting. The reasons are twofold: (1) The shape of a time series may be too complex for a basic mathematical expression to fit. It is better to divide the time series into multiple non-overlapped intervals and fit them one by one. (2) The shorter the interval is, the simpler the expression is. Sometimes we are more interested in the shape of a specific interval, and a simple expression without higher terms is easier to understand. In the top middle panel, a total number of 25 fitting expressions are provided. The tool can choose the best one automatically based on the metric of RMSE. In the middle panel, the parameters are displayed to be adjusted. The right panel displays the fitting results as text, and the bottom panel visualizes the fitting results.

Unlike the reproduction method, even if the original workload consists of $N_S$ sessions, the number of $S$ extracted from the relational model is not necessarily $N_S$ in the fitting method because the shape of the original intensity has been changed by the decomposition algorithm.

\subsubsection{Generation}

In the generation method, we create the new intensity from scratch based on self-specified characteristics. In this way, we can customize the shape of the simulated workload. For different known parameters and required characteristics of the new intensity, two existing methods of time series generation, LIMBO \citep{kistowski2017modeling} and TSAGen \citep{wang2021tsagen}, can be integrated into the generation method as follows.

\begin{enumerate}[label={\textbf{\arabic*)}},leftmargin=0cm,itemindent=1cm,labelwidth=.5cm]
    
    \item{\textbf{LIMBO-based Method:} We follow LIMBO which allows a DSL-based workload intensity definition and make some changes for better intervention. The workload intensity is represented with the trend component $trend$, the season component $season$, the burst component $burst$, and the noise component $noise$. The additive combinator is applied to merge above components:
    
    \begin{equation}
        I(t)=trend(t)+season(t)+burst(t)+noise(t)
    \end{equation}
    
    Table~\ref{Details of the generation method based on LIMBO} shows the details of each component with related parameters and the generation process where $mod$ means the modulo operation. The noise component is generated by sampling from the Gaussian distribution $N(\frac{\eta_9 + \eta_8}{2},\frac{\eta_9 - \eta_8}{6})$ where 99.7\% of data occurs within three standard deviations of the mean constrained by the minimum value $\eta_8$ and the maximum value $\eta_9$. The other three components are generated by interpolation functions such as polynomials. The length of the trend component is in multiples of seasonal iterations which is calculated by $\eta_1 \times \eta_2$. Since the determination of nonlinear interpolation functions generally requires three points, three representative points, the starting point, the endpoint, and the midpoint whose values are $c_1$, $c_2$, and $c_3$ respectively are selected for the interpolation of $trend$. The generation of the seasonal component is shown in Figure~\ref{LIMBOSeason}. Given the first peak $c_6$, the lask peak $c_7$, and the number of peaks $\eta_3$, additional peaks are derived from linear interpolation which is the most intuitive. The distance between the starting and the first peak is assumed to be equal to that between the last peak and the end, denoted by $0.5 \times \eta_m$. In the boundary interval between the starting and the first trough (or the last trough and the end), $season$ is generated by interpolation of the boundary point, the peak, and the trough. In the intervals between two adjacent troughs, $season$ is generated by interpolation of the peak and troughs. The burst component defines recurring bursts under the assumptions of the same recurring interval $\eta_6$ and the axisymmetric burst. Only the point in the burst has the non-zero value which is also generated by the interpolation function given the burst width $\eta_7$ and the burst peak $c_8$ (the starting and the end value of a burst is 0). Besides, the first burst peak offset $\eta_5$ controls the offset and is under the assumption of $\eta_5< \eta_6$.

    \begin{table*}[!t]
        \caption{Details of the generation method based on LIMBO.}
        \footnotesize
        \label{Details of the generation method based on LIMBO}
        \centering
        \setlength{\tabcolsep}{3mm}{
            \begin{tabular}{c|c|c|c}
                \hline
                Component                  & Parameter   & Description         & Generation \\
                \hline
                \multirow{6}{*}{$trend$}   & $\eta_1$    & number of seasonal periods in one trend      & \multirow{6}{*}{$trend(t) = g_1(t-t_s,\eta_1 \times \eta_2,c_1,c_2,c_3)$}            \\
                \cline{2-3}                & $\eta_2$    & seasonal period (same as that in $season$)              &            \\
                \cline{2-3}                & $c_1$       & starting point value     &            \\
                \cline{2-3}                & $c_2$       & endpoint value     &            \\
                \cline{2-3}                & $c_3$       & midpoint value     &            \\
                \cline{2-3}                & $g_1$       & interpolation function for trend shape      &            \\                
                \hline
                \multirow{8}{*}{$season$}  & $\eta_2$    & period              & \multirow{5}{*}{$season(t)=\left\{\begin{array}{l}
                    g_2(t_c,\eta_s,c_4,c_5,c_6), 0\le t_c < \eta_s \\ 
                    g_2(t_c,\eta_r,c_5,c_5,c_6+\frac{(i+1)(c_7-c_6)}{\eta_3-1} ), \\ \quad\  i \times \eta_r \le t_c-\eta_s < (i+1) \times \eta_r \\ 
                    g_2(t_c,\eta_s,c_5,c_4,c_7), \eta_2 - \eta_s \le t_c \le \eta_2 \\
                \end{array}\right.$}               \\
                \cline{2-3}                & $\eta_3$    & number of peaks     & \\
                \cline{2-3}                & $\eta_4$    & interval between first and last peaks         &                                             \\
                \cline{2-3}                & $c_4$       & starting and end value of one iteration     &                                             \\
                \cline{2-3}                & $c_5$       & trough between two peaks  &                                           \\
                \cline{2-3}                & $c_6$       & first peak  & where $t_c=(t-t_s)\; mod\;  \eta_2$, $\eta_s=0.5 \times (\eta_m+\eta_r)$ ,\\
                \cline{2-3}                & $c_7$       & last peak     &\multirow{2}{*}{$\eta_m=\eta_2-\eta_4$, $\eta_r=\eta_4/(\eta_3-1)$ and $i < \eta_3-2$, $i \in \mathbb{N}$}           \\
                \cline{2-3}                & $g_{2}$     & interpolation function for seasonal shape     &           \\
                \hline
                \multirow{5}{*}{$burst$}                & $\eta_5$    & first burst peak offset                & \\
                \cline{2-3}                & $\eta_6$    & interval between two bursts  &      \multirow{2}{*}{$burst(t)=\left\{\begin{array}{l}
                    g_3(t_c,\eta_7,0,0,c_8), 0\le t_c\le \eta_7 \\ 
                    0, t_c>\eta_7
                \end{array}\right.$}      \\
                \cline{2-3}                & $\eta_7$    & burst width         &            \\
                \cline{2-3}                & $c_8$       & burst peak          & \multirow{2}{*}{where $t_c=(t-t_s-\eta_5+0.5\times \eta_7)\; mod\; \eta_6$}           \\
                \cline{2-3}                & {$g_3$}       & interpolation function for burst shape           &            \\
                \hline
                \multirow{2}{*}{$noise$}   & $\eta_8$    & minimum value                & \multirow{2}{*}{$noise(t) = Normal(\frac{\eta_9 + \eta_8}{2}, \frac{\eta_9 - \eta_8}{6})$}                                                                                 \\
                \cline{2-3}                & $\eta_9$    & maximum value  &            \\
                \hline
            \end{tabular}
        }
        \vspace{-4mm}
    \end{table*}
    
    \begin{figure}[!t]
        \centering
        \includegraphics[width=0.48\textwidth]{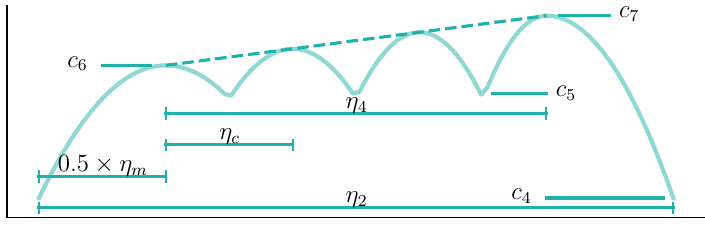}
        \caption{Seasonal component of the LIMBO generation method.}
        \label{LIMBOSeason}
        \vspace{-4mm}
    \end{figure}
    
    } 
    
    \item{\textbf{TSAGen-based Method:} TSAGen is a time series generation tool for controllable and anomalous key performance indicator data generation. For the property in the task of workload simulation, only the normal data generation stage is integrated into workload intensity modeling. To be specific, the workload intensity is represented with the trend component $trend$, the seasonal component $season$, and the noise component $noise$ of the same length:
    
    \begin{equation}
        I(t)=trend(t)+season(t)+noise(t)
    \end{equation}
    
    \begin{table*}[!t]
        \caption{Details of the generation method based on TSAGen.}
        \label{Details of the generation method based on TSAGen}
        \footnotesize
        \centering
        \setlength{\tabcolsep}{3mm}{
            \begin{tabular}{c|c|c|c}
                \hline
                Component                  & Parameter   & Description         & Generation \\
                \hline
                \multirow{2}{*}{$trend$}   & $\theta_1$  & level of trend      & \multirow{2}{*}{$trend(t) = \theta_1 + \theta_2 \times (t-t_s)$}            \\
                \cline{2-3}                & $\theta_2$  & slope of trend      &            \\
                \hline
                \multirow{7}{*}{$season$}  & $\theta_3$  & amplitude           &            \\
                \cline{2-3}                & $\theta_4$  & frequency           & \multirow{2}{*}{$season(t)=(\sum_{i=1}^{\theta_5}\bigoplus(\theta_3\times a_i)cycle_{i,f_i*1/\theta_4})_{t-t_s},$}                                    \\
                \cline{2-3}                & $\theta_5$  & number of cycle     &                                             \\
                \cline{2-3}                & $k_1$       & drift degree of amplitude  & where $a_i\sim U(1,1+k_1), k_1\in[0,+\infty ),$                                            \\
                \cline{2-3}                & $k_2$       & drift degree of frequency  & $f_i\sim U(1,1+k_2), k_2\in[0,+\infty ),$                                                                                      \\
                \cline{2-3}                & $d_1$       & recursion depth     & and $cycle_{i,j}=sample_i(F_{RMDF}(d_1, d_2), j)$           \\
                \cline{2-3}                & $d_2$       & forking depth       &            \\
                \hline
                \multirow{3}{*}{$noise$}   & $\theta_6$  & skewness            & \multirow{3}{*}{$noise(t) = Gamma(\theta_6, \theta_7, \theta_8)$}                                                                                 \\
                \cline{2-3}                & $\theta_7$  & location            &            \\
                \cline{2-3}                & $\theta_8$  & scale               &            \\
                \hline
            \end{tabular}
        }
        \vspace{-4mm}
    \end{table*}
    
    Details of each component with related parameters and the generation process are shown in Table~\ref{Details of the generation method based on TSAGen}. The trend component is generated by a linear function by setting the level and the scope. The seasonal component is generated by connecting cycles whose shape is modeled by the Random Midpoint Displacement Fractal (RMDF) algorithm~\citep{fournier1982computer}. $\sum \bigoplus$ means multiple cycles are concatenated in order. $a_i$ and $f_i$ are the drift factors of amplitude and frequency sampled from the uniform distributions $U(1, 1+k_1)$ and $U(1, 1+k_2)$ respectively for representing drifts such as holiday effects. $cycle_{i,f_i*1/\theta_4}$ denotes the $i$-th cycle of the length $f_i*1/\theta_4$ generated by sampling (marked as $sample_i$) on the function $F_{RMDF}$ obtained from the RMDF algorithm. Figure~\ref{The rmdf process} displays the RMDF process. Given two initial points $P_1(0,0),P_2(1,0)$, the control point in the perpendicular bisector is chosen. For a relatively smooth shape, the distance between the control point and the midpoint is sampled from the Gaussian distribution $N(0,0.25)$ and decreases by half each recursion. Through trigonometric functions and coordinate operations, the function of the curve which connects the initial points and the control point is generated (refer to the blue lines in Figure~\ref{The rmdf process} for an example). In each new interval generated by adjacent initial points and control points, the process is repeated recursively which is controlled by the recursion depth $d$ (refer to the purple and yellow lines in Figure~\ref{The rmdf process} for an example). For small differences in shape modeling, a base function sharing by all cycles is generated by setting the certain recursion depth $d_1$. Then $d_2$ recursions are applied in the base function for each cycle. Based on parameters $\theta_3,\theta_4,\theta_5,k_1,k_2$ controlling characteristics and $d_1,d_2$ controlling shapes, the seasonal component is generated where the value at the timestamp $t$ is the $(t-t_s)$-th element. The Pearson type \uppercase\expandafter{\romannumeral3} distribution, also known as the Gamma distribution, is used to generate the noise component due to the nonnegativity of Gamma-distributed random variables and the controllable statistical properties of the Gamma distribution. Given the skewness $\theta_6$, the location $\theta_7$, and the scale $\theta_8$, the probability density function can be calculated. Then the noise component is generated based on the probability density function.
    
    \begin{figure}[!t]
        \centering
        \includegraphics[width=0.5\textwidth]{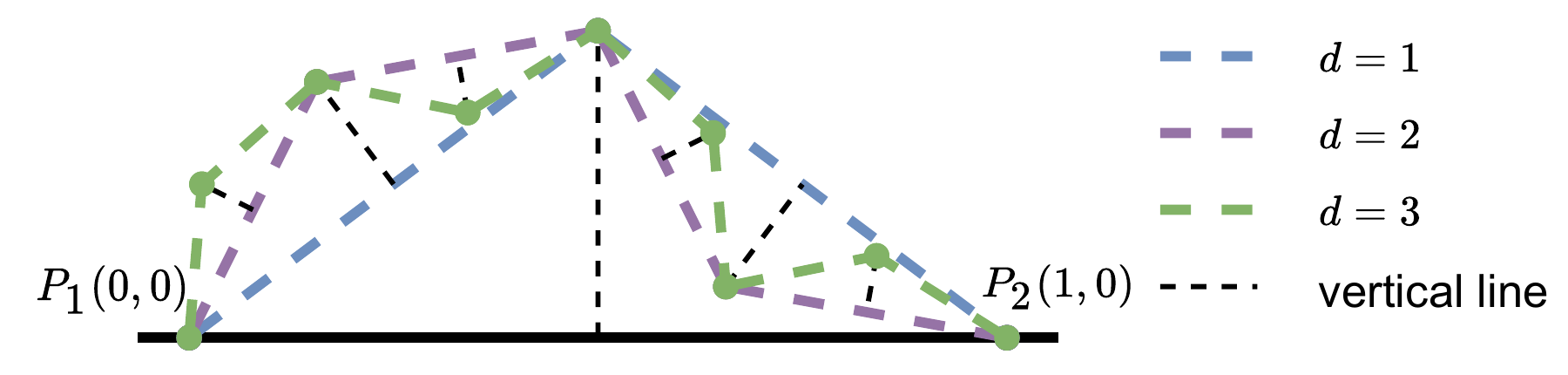}
        \caption{The RMDF process. The black dotted line represents the perpendicular bisector. Blue, purple and yellow lines represent curves generated by setting $d=1,2,3$ in RMDF respectively.}
        \label{The rmdf process}
        \vspace{-4mm}
    \end{figure}
    }
    
\end{enumerate}

Similar to the fitting method, the number of $S$ extracted from the relational model depends on the generated intensity rather than the original intensity.

\subsubsection{Think Time}

Not only is the time attribute of a session characterized by the start time, but also by think times between consecutive user behaviors. The aforementioned three methods have modeled the workload intensity in terms of the session start time. Transitions between consecutive user behaviors within a session should also be annotated with the think time. For $N_u$ types of user behaviors, think times in each session can be denoted by a $N_u \times N_u$ matrix $\boldsymbol{T}$ where the element $T_{r,c}$ in the $r^\text{th}$ row and the $c^\text{th}$ column denotes the think time from the user behavior $u_r$ to the user behavior $u_c$. To simplify the problem, we assume think times of different user behavior pairs share the same distribution and all think times can be denoted as a random variable $\tau$. To model the distribution of think times, we use the non-parametric kernel density estimator (KDE) ${\widehat f}_h(\tau)$:

\begin{equation}
    {\widehat f}_h(\tau)=\frac1{n_{tt}}{\textstyle\sum_{i=1}^{n_{tt}}}\phi_h\left(t-t_i\right)=\frac1{n_{tt}h}{\textstyle\sum_{i=1}^{n_{tt}}}\phi(\frac{t-t_i}h)
\end{equation}

\noindent where $n_{tt}$ is the number of think times (or the number of two consecutive requests) in the original workload, $h$ is the bandwidth and $\phi$ is the kernel. We define $\phi$ as the standard normal density $\phi(\tau) \sim N(0,1)$:

\begin{equation}
    \phi\left(\tau\right)=\frac1{\sqrt{2\mathrm\pi}}e^{-\frac12t^2}
\end{equation}

To find the ideal bandwidth $h$, the rule-of-thumb approach \citep{KDE} is applied by estimating standard deviation $\sigma_{tt}$ from the think time in the original workload:

\begin{equation}\label{KDE bandwidth estimation}
    h=1.06\sigma_{tt}{n_{tt}}^{-1/5}
\end{equation}

\subsection{Simulated Workload Generation}

With the preliminaries of user behavior abstraction and workload intensity modeling, we can generate the simulated workload. Given user groups $\{C_1, C_2, \cdots , C_{N_c}\}$ identified in Section \ref{User grouping} and the corresponding numbers of user behavior sequences $\{\hat{N}_1, \hat{N}_2, \cdots , \hat{N}_{{N_c}}\}$, the probability of each simulated user behavior sequence belonging to the $C_i$ cluster is $\hat{N}_i / \sum_{j=1}^{N_c}\hat{N}_j$. After determining the user group, we can extract the user behavior sequence $S$ from the relational model of the group using the idea of Depth-First Search (DFS). We start traversal from the dummy node $u_s$ and look for its adjacent nodes. According to the transition probability, we move to the next node and continue this loop until the dummy node $u_e$ is selected. We do not mark any node as visited in the search, so it is possible that a node may be visited repeatedly when there exists a cycle. This strategy guarantees the diversity of generated user behavior sequences. For each $S$, we need to perform a DFS, and the value of $I(t)$ decides how many $S$ we need to extract for the bucket containing the timestamp of $t$.

For the $i^\text{th}$ bucket which represents the time span of $[t_s+(i-1) \times \delta, t_s+i \times \delta)$, assume there are $k_i$ sessions that belong to this bucket in the simulated workload. If all $k_i$ sessions start at the timestamp of $t_s+(i-1) \times \delta$, the user concurrency will be skewed to the left point of each interval especially when $\delta$ is set to a large value or $I(t_s+(i-1) \times \delta)$ is too large. To ensure the stability of $\mathcal{T}$, we suggest distributing each session evenly in the bucket. Let $\triangle t_i = \delta / k_i$. Then, in the $i$-th bucket, the first session starts at $t_s+(i-1) \times \delta$, the second session starts at $t_s+(i-1) \times \delta + \triangle t_i$ and so on. The duration of each session is not constant because of the diversity of $S$ and the probabilistic think time, but we do not need to pay attention to when a session ends. We also design a domain-specific language (DSL) to describe the workload features in a human-readable way. Non-experts just need to submit a file following the grammar rules for quick execution of LWS. For more details, please refer to supplemental materials in Section \ref{Data availability}.

\section{Case Study}\label{sec:case study}

In this section, we apply the proposed framework LWS to a session-based application and aim to address the following research questions:

\begin{itemize}

 \item{\textbf{RQ1:}} How accurately does the simulated workload generated by LWS match the original workload?

 \item{\textbf{RQ2:}} How effective is the intervention of the simulated workload generated by LWS?
 
 \item{\textbf{RQ3:}} How do different workload intensity modeling methods impact the effectiveness of the intervention?

\end{itemize}

\subsection{Experiment Setup}

\subsubsection{Benchmark System}\label{Benchmark system}

Our studies are conducted in an open-source cloud-native microservices demo application, Hipstershop (also known as Online Boutique)\footnote{https://github.com/GoogleCloudPlatform/microservices-demo}. It is a typical web-based e-commerce application widely used in research on the microservice operation. It consists of 10 microservices written in Java, Python, Go, Node.js, and C\#, and provides an extra microservice for original workload generation. For each user session, a unique session ID will be generated automatically and recorded in logs. We deploy Hipstershop in an open-source cloud-native system Kubernetes\footnote{https://kubernetes.io/} with 5 physical machines and 4 virtual machines. Monitoring tools including Elasticsearch, Filebeat, Prometheus\footnote{https://prometheus.io/} and Jaeger\footnote{https://www.jaegertracing.io/} are also deployed in the Kubernetes for monitoring data collection. We implement LWS with Python and deploy the application on another physical server. The details of the deployment environment are shown in Table~\ref{Details of the deployment environment}. We have also summarized the detailed resource recommendations and requirements, as well as the related deployment processes and files. For more details, please refer to supplemental materials in Section \ref{Data availability}.

\begin{table}[!t]
    \caption{Details of the deployment environment.}
    \label{Details of the deployment environment}
    \footnotesize
    \centering
    \setlength{\tabcolsep}{3mm}{
        \begin{tabular}{c|c|c}
            \hline
            System                  & Component        & Detail \\
            \hline
            \multirow{11}{*}{Kubernetes}   & \multirow{2}{*}{1 physical master}   & 64-core 2.30GHz CPU,           \\
                                          &                                      & 64GB RAM       \\
            \cline{2-3}                   & \multirow{2}{*}{4 physical nodes}    & 4-core 2.66GHz CPU,             \\
                                          &                                      & 8GB RAM       \\
            \cline{2-3}                   & \multirow{2}{*}{4 virtual nodes}     & 4-core 2.10GHz CPU,            \\
                                          &                                      & 16GB RAM       \\
            \cline{2-3}                   & \multirow{2}{*}{Hipstershop}         & 4 pod replicas for    \\
                                          &                                      & each microservice       \\
            \cline{2-3}                   & Elasticsearch       & version 7.13.2           \\
            \cline{2-3}                   & Filebeat            & version 7.13.2           \\
            \cline{2-3}                   & Prometheus          & version 2.10.0            \\
            \cline{2-3}                   & Jaeger              & version 1.28           \\
            \hline
            \multirow{3}{*}{LWS Driver}   & \multirow{2}{*}{1 physical server}   & 64-core 2.40GHz CPU,           \\
                                          &                                      & 64GB RAM       \\
            \cline{2-3}                   & Python              & version 3.6.9           \\
            \hline
        \end{tabular}
    }
    \vspace{-4mm}
\end{table}

\subsubsection{Dataset}\label{dataset}

We use two datasets $\mathcal{A}$ and $\mathcal{B}$ as subjects where $\mathcal{A}$ is the well-designed original workload and $\mathcal{B}$ is the real-world workload in the production system:

\begin{enumerate}[label={\textbf{\arabic*)}},leftmargin=0cm,itemindent=1cm,labelwidth=.5cm]

 \item{\textbf{Dataset} $\mathcal{A}$\textbf{:} We use K6\footnote{https://k6.io/}, an open-source load testing tool, to simulate a real-world load on Hipstershop as the original workload. The original user behavior sequence $S$ consists of 7 to 20 user behaviors along with 0 to 11 automatic redirections, and the think time $\tau$ is generated independently and identically from $\phi(\tau) \sim N(5,2)$. To avoid the negative and exceptional value, the think time will be resampled from $\phi(\tau) \sim N(5,2)$ until it falls within the interval $[1,15]$. Both the K6 script running for 3 hours and 40 minutes in total and the scenario with excellent statistical properties such as multiple peaks, obvious seasonality and trend changes which are intuitive and common in real production workloads~\citep{feng2022fast} have been applied in 2022 CCF AIOps Challenge\footnote{https://www.bizseer.com/index.php?m=content\&c=index\&a=\\show\&catid=25\&id=83}.} The number of virtual users ramps up from 0 to 60 in 10 minutes, keeps 60 for 40 minutes, ramps down from 60 to 20 in 10 minutes, keeps 20 for 20 minutes, ramps up from 20 to 80 in 10 minutes, keeps 80 for 40 minutes, ramps down from 80 to 20 in 10 minutes, keeps 20 for 20 minutes, ramps up from 20 to 60 in 10 minutes, keeps 60 for 40 minutes and ramps down from 60 to 0 in 10 minutes. The user behavior sequence $S$ will not be forcefully interrupted at the boundary of the test duration and the user number changes. In case of skewed metrics and unexpected test results, every $u \in S$ will be executed and K6 will wait for any iterations in progress to finish. Hence, the actual execution time will be little greater than 3 hours and 40 minutes.

 \item{\textbf{Dataset} $\mathcal{B}$\textbf{:} The dataset $\mathcal{B}$ is generated based on \textbf{UserBehavior}\footnote{https://tianchi.aliyun.com/dataset/dataDetail?dataId=649} which is a subset of Taobao user behavior data containing millions of user-item interactions during November 25 to December 03, 2017, offered by Alibaba. Due to the differences in business functions and capacities between the production system of \textbf{UserBehavior} and Hipstershop, it is impossible to reproduce the exact workload of \textbf{UserBehavior} based on Hipstershop. Therefore, we refer to the workload intensity characteristics of \textbf{UserBehavior} as the target of the intervention rather than the complete reproduction. To generate the real-world original workload intensity within the capacity of the benchmark system in Section~\ref{Benchmark system}, we count the number of new concurrent users per 10 minutes from November 25 to December 03, 2017 (in UTC+8 Zone), multiply each number by 0.02 and round results down for adapting to the load capacity in our benchmark system. We adopt the same strategy in the user behavior and think time in $\mathcal{A}$ and reproduce the workload as the dataset $\mathcal{B}$ as the approximated original workload.}

\end{enumerate}

\subsubsection{Evaluation Metric}

To evaluate the similarity between the original workload and the simulated workload quantitatively, we introduce two categories of metrics: 

\begin{enumerate}[label={\textbf{\arabic*)}},leftmargin=0cm,itemindent=1cm,labelwidth=.5cm]

 \item{\textbf{Static Metrics:} Static metrics are obtained from statistics of generated user behavior sequences and thus they will not be affected by the execution of HTTP requests in the SUT. Two workload characteristics, the session length distribution and the behavior distribution (including the relative invocation frequency of different request types and the number of distinct session types) are applied in static metrics. Regardless of dynamic factors such as the request time consuming, the session length distribution is determined by two important factors: the number of requests per session and think time. For the number of requests per session, we refer to the session metrics used in WESSBAS. The minimum value (denoted by ``$Min$"), the maximum value (denoted by ``$Max$"), the $25^\text{th}$ quartile (denoted by ``$Q1$"), the median value (denoted by ``$Q2$"), the $75^\text{th}$ quartile (denoted by ``$Q3$"), the mean value (denoted by ``$Mean$") and the 0.95 confidence interval (denoted by ``${CI}_{0.95}$") are computed for analysis. For the think time, we use the metric Intersection over Union ($IoU$) which quantifies the degree of overlap between the original think time distribution and the estimated think time distribution for evaluation. For the behavior distribution, we refer to the request count statistics used in WESSBAS. The relative invocation frequency of each request type and the number of distinct session types (denoted by ``${N}_{ds}$") are computed. Besides, we record the time cost of user behavior abstraction (denoted by $AbstractionTimeCost$) for the evaluation of efficiency.}
 
 \item{\textbf{Dynamic Metrics:} Dynamic metrics are computed by monitoring metrics such as CPU utilization and memory usage in the form of time series collected from the SUT under the workload. Note that the investigation of dynamic metrics in workload simulation is proposed in one of our submitted papers \citep{workloadsimulationevaluation}, and thus we directly use its methodology in this work. Monitoring metrics are grouped into two categories by the degree of the correlation with business due to their extreme large quantity and variety (for more details, please refer to \cite{workloadsimulationevaluation}) and the linked dataset in Section \ref{Data availability}. In each category, the Extend-SBD based on the Shape-based distance (SBD) \citep{paparrizos2015k} which describes both shape and intensity similarities given the weighting factor $\alpha$ is measured for each metric. Then the weighted average $distance \in [0,2]$ of the Extend-SBD values in each category is calculated. Lower $distance$ shows the higher similarity between workloads. The weighted average calculated from metrics weakly correlated with business is denoted by $ESBD(weak)$, and the weighted average calculated from metrics strongly correlated with business is denoted by $ESBD(strong)$. Given the similarity threshold $\mu$, the normalized similarity score $score \in [0,1]$ is calculated:
 
 \begin{equation}
    score = \frac{\mu}{\mu + distance}
 \end{equation}
 
 For $ESBD(weak)$ and $ESBD(strong)$, the normalized similarity scores are calculated and denoted by $score_w$ and $score_s$ respectively. Higher $score$ shows a higher similarity between workloads. The composite score denoted by $score_c$ is calculated by the weighted summation where $\beta$ is a hyperparameter:
 
 \begin{equation}
    score_c = (1-\beta)score_w + \beta score_s, \beta \in {(}0,1{]}
 \end{equation}
 
 We set $\alpha=0.5$, $\mu=1.0$, $\beta=0.9$ in the following evaluation based on dynamic metrics.
 
 } 

\end{enumerate}

\subsection{RQ1: Accuracy of Simulated Workload}

The evaluation of simulation accuracy (RQ1) depends on the comparison between the original workload (denoted by \textit{OW}) in the dataset $\mathcal{A}$ and the simulated workload. The output logs and monitoring metrics under the original workload are collected. Now, we have to assume that we have no idea of the original workload which is specified by the K6 script and try to simulate it via these logs.

\subsubsection{Log Collection and Transformation}

\begin{figure}[!t]
\centering\includegraphics[width=0.5\textwidth]{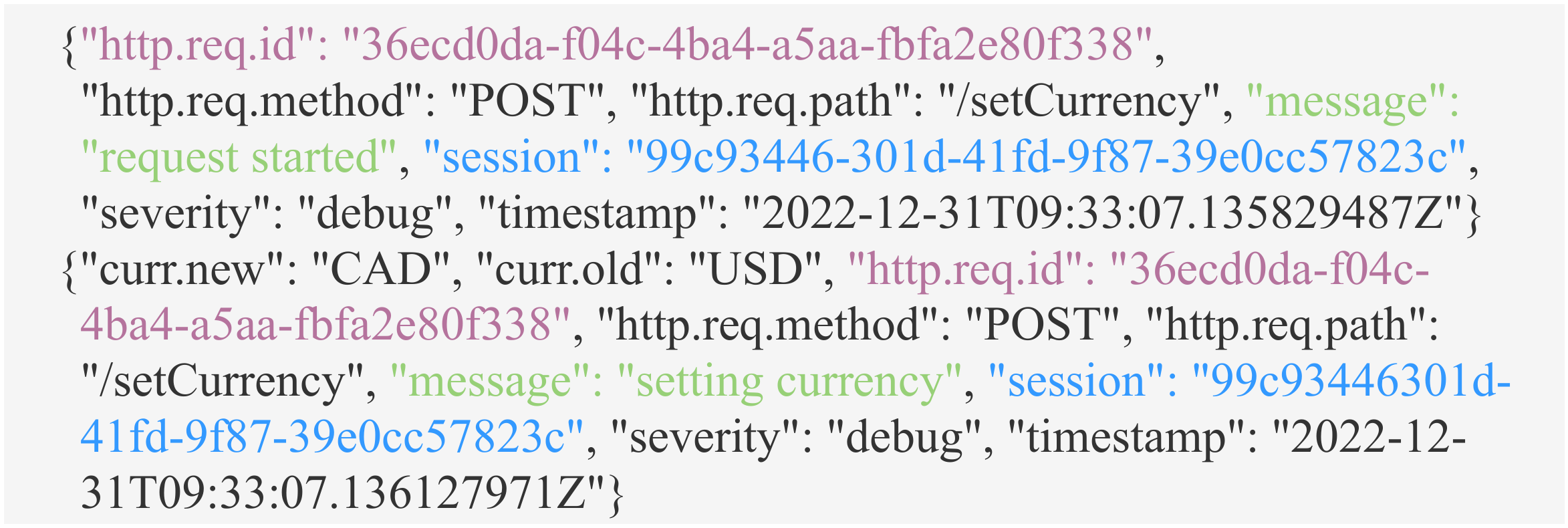}
\caption{An example of two consecutive HTTP request logs.}\label{HTTP request logs}
\vspace{-4mm}
\end{figure}

Using the regex match, we can remove the redundant logs and only keep the HTTP request logs. After that, we group these logs by the field of ``session" and sort the logs within each session by the field of ``timestamp". Figure~\ref{HTTP request logs} illustrates an example of two consecutive HTTP request logs. We can find some useful patterns: (1) The ``http.req.id" field is the unique identifier to group the HTTP requests within a session. (2) Each HTTP group follows a common template where the value of the ``message" field is ``request started" in the first line and ``request complete" in the last line. (3) The ``message" field in the middle line indicates a certain type of user behavior, and there are 6 different types of user behaviors in total including ``home" ``setting currency" ``serving product page" ``view user cart" ``adding to cart" ``placing order". Note that ``placing order" is always followed by ``order placed" in an HTTP group and they refer to the same HTTP request, so we use ``placing order" as the name of the user behavior for simplicity. A possible reason for the unnecessary alias in these logs is bad coding practice. At last, a total number of 7013 sessions can be obtained.

\subsubsection{User Behavior Abstraction}

In addition to the methodology in Section \ref{sec:user behavior abstraction} (denoted by \textit{SW (HAC/Markov)}, cluster num=3), we perform the following different user behavior abstraction methods for comparison:

\begin{itemize}
    \item \textit{SW (Monolithic)}: Its user behavior abstraction is based on a monolithic Markov probabilistic transition matrix without clustering and a relational model with Synoptic invariants. 
    
    \item \textit{SW (HAC/SGT)}: Its user behavior abstraction is based on the Sequence Graph Transform embedding \citep{ranjan2022sequence} with the same cluster and relational model construction strategy in Section \ref{sec:user behavior abstraction} (cluster num=3). 

    \item \textit{WESSBAS}: Its user behavior abstraction is based on the Markov probabilistic transition matrix, the X-means cluster algorithm (the cluster num found is 4), and the relational model with proposed invariants as introduced in \cite{vogele2018wessbas}.
    
\end{itemize}

\begin{figure}[!t]
\centering
\includegraphics[width=0.5\textwidth]{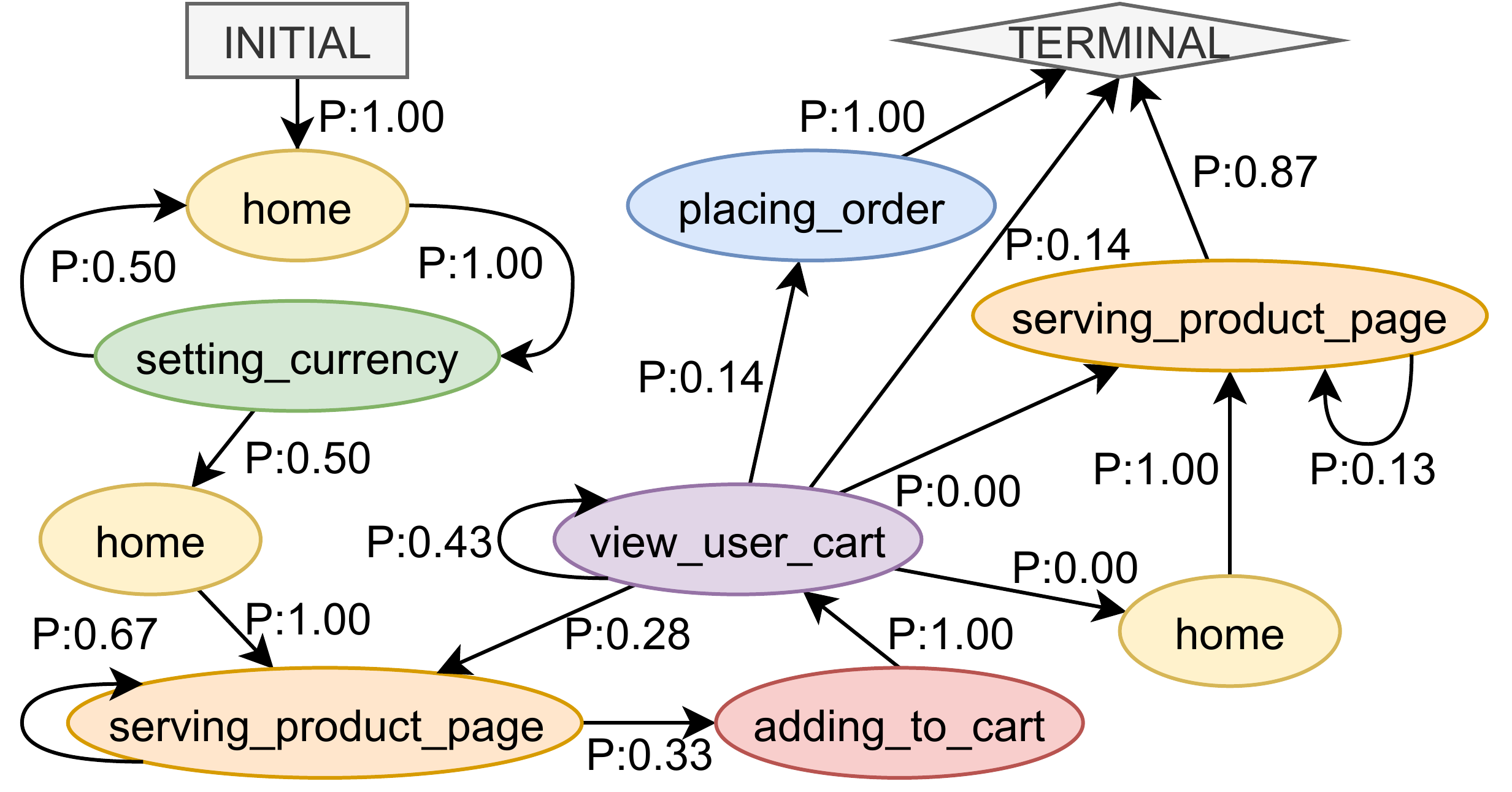}
\caption{The relational model of user group 1 in SW (HAC/Markov) constructed on $\mathcal{A}$.}
\label{The abstraction of user behaviors}
\vspace{-4mm}
\end{figure}

The resulting user behavior abstraction of user group 1 in \textit{SW (HAC/Markov)} is illustrated in Figure~\ref{The abstraction of user behaviors} as an example. $P_{a, b}$ denotes the transition probability from the user behavior $u_a$ to the user behavior $u_b$. It can be seen from the figure that $P_{a, b}$ equals 0.00 or 1.00 sometimes. If a transition probability has a value less than 0.01, it will be approximated by 0.00. There are two situations that result in a transition probability of 1.00:

\begin{itemize}
    \item One-to-one Transition: For example, in the original workload, the first user behavior is always ``home". Therefore, the transition probability from ``initial" to ``home" is 1.00.
    
    \item Redirection: For example, the reason for the transition probability of 1.00 from ``setting\_currency" to ``home" is that the HTTP request will be redirected to $u_b$ automatically after $u_a$ has been executed. We recommend manually analyzing the frontend code, or sending the HTTP request for $u_a$ to the SUT and then analyzing the output logs or the HTTP status code (such as 302) to identify the redirection operation and avoid redundant user behaviors.
    
\end{itemize}

\subsubsection{Workload Intensity Modeling}

\begin{figure}[!t]
\centering
\includegraphics[width=0.46\textwidth]{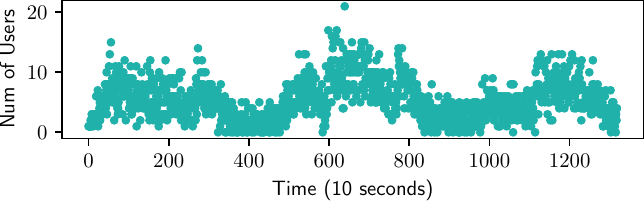}
\caption{The intensity via \textit{Reproduction} of $\mathcal{A}$ setting $\delta=10$.}
\label{Reproduction intensity density of A}
\end{figure}

We use \textit{Reproduction} which contains no intervention when simulating the workload of $\mathcal{A}$. Figure~\ref{Reproduction intensity density of A} displays the intensity scatter setting $\delta=10$. The shape of the workload intensity generated by the reproduction method is intuitively similar to that of the number of virtual users in the original workload. However, the value of the point in the intensity is significantly smaller than the number of virtual users in the original workload. The reason is that the number of users set by K6 describes the concurrency rather than the increment. In other words, K6 will count a user in the concurrency until behaviors belonging to the user finish. Different from K6, each point in the generated intensity describes the increment indicating how many additional users will start their behaviors. Besides, the points in the intensity are intuitively even, which ensures the stability of the simulated workload.

\begin{figure}[!t]
\centering
\includegraphics[width=0.48\textwidth]{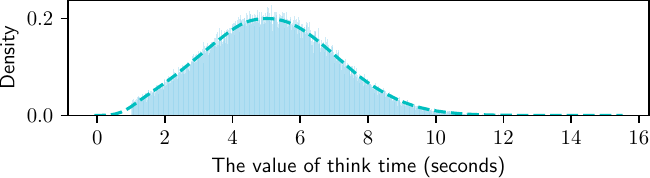}
\caption{The modeled probability density function of think times in SW (HAC/Markov) constructed on $\mathcal{A}$.}\label{The density function of the think time}
\vspace{-4mm}
\end{figure}

For the think time, we calculate the time difference between the timestamp recorded in the ``request complete'' log of each behavior and the timestamp recorded in the ``request start'' log of its next behavior as an approximation. Then we set the bandwidth calculated by Equation~\ref{KDE bandwidth estimation} and Figure~\ref{The density function of the think time} displays the estimation result. The dashed line shows the probability density function modeled by KDE and the light blue histogram shows the original distribution of the think time. It can be seen from Figure~\ref{The density function of the think time} that the modeled probability density function can intuitively fit the distribution of the think time well although there exist interval limitations in the original distribution.

\subsubsection{Simulated Workload Generation}

We sample 7013 user behavior sequences from the user abstraction model and schedule them by the generated workload intensity. For the simulated workload, the user behavior sequence in each session differs from each other and we need to send each HTTP request in an exact timestamp. Traditional load testing tools which only focus on the number of concurrent users conveniently are not suitable for this task. For example, K6 executes a group of sessions simultaneously from a high level while lacking the fine-grained control of the start time for each HTTP request. As a result, we develop a useful script based on multithread scheduling to execute the simulated workload. The whole process starting from monitoring the application until the simulated workload is acquired is around 9 hours, including the original workload (about 3 hours and 40 minutes), collection and transformation of logs and monitoring metrics (about 1 hour with another 20 minutes waiting for stability in monitoring tools), LWS extraction (about 10 minutes), and the simulated workload (about 3 hours and 40 minutes). Most of the time cost is in the workload execution and data preparation.

\subsubsection{Quantitative Analysis of Accuracy}

\begin{table*}[!t]
\caption{Static metrics in terms of the session length distribution and the time cost of user behavior abstraction between the original workload and the simulated workload in $\mathcal{A}$.}
\label{RQ1 session length static metrics}
\footnotesize
\begin{center}
\begin{tabular}{cccccccccc}
\hline
Workload Type   & $Min$ & $Q1$ & $Q2$ & $Q3$ & $Max$ & $Mean$ & ${CI}_{0.95}$     & $IoU$   & $AbstractionTimeCost$ \\ \hline
\textit{OW}              & 7     & 9    & 19   & 24   & 31    & 18.74  & {[}18.56,18.91{]} & -       & -        \\
\textit{SW (Monolithic)} & 4     & 11   & 16   & 24   & 100   & 19.29  & {[}19.00,19.58{]} & 73.25\% & 12.06s   \\ 
\textit{SW (HAC/Markov)} & 4     & 9    & 14   & 23   & 155   & 18.77  & {[}18.41,19.12{]} & 73.46\% & 26.99s   \\ 
\textit{SW (HAC/SGT)}    & 4     & 9    & 15   & 23   & 142   & 18.19  & {[}17.88,18.49{]} & 73.71\% & 39.27s   \\ 
\textit{WESSBAS}         & 4     & 10   & 14   & 21   & 93    & 16.57  & {[}16.33,16.80{]} & 73.45\% & 24.34s   \\ 
\hline
\end{tabular}
\end{center}
\vspace{-4mm}
\end{table*}

\begin{table*}[!t]
\caption{Static metrics in terms of the behavior distribution between the original workload and the simulated workload in $\mathcal{A}$.} 
\label{RQ1 behavior length static metrics}
\footnotesize
\begin{center}
\begin{tabular}{cccccccc}
\hline
Workload Type      & adding\_to\_cart & home    & placing\_order & serving\_product\_page & setting\_currency & view\_user\_cart & $N_{ds}$      \\ \hline
\textit{OW}                 &  9.95\%          & 18.05\% & 2.98\%         & 39.44\%                & 10.67\%           & 18.92\%          & 131415 \\
\textit{SW (Monolithic)}    &  9.82\%          & 17.72\% & 3.01\%         & 40.03\%                & 10.47\%           & 18.95\%          & 135262 \\
\textit{SW (HAC/Markov)}    &  9.81\%          & 18.11\% & 2.97\%         & 39.85\%                & 10.66\%           & 18.61\%          & 131607 \\
\textit{SW (HAC/SGT)}       &  9.78\%          & 18.40\% & 2.97\%         & 39.60\%                & 10.95\%           & 18.30\%          & 127536 \\
\textit{WESSBAS}            & 10.10\%          & 18.72\% & 2.74\%         & 37.88\%                & 11.89\%           & 18.67\%          & 116179 \\
\hline
\end{tabular}
\end{center}
\vspace{-4mm}
\end{table*}

We compare the original workload and the simulated workloads in the dataset $\mathcal{A}$ quantitatively by static metrics and dynamic metrics after simulated workload generation. Static metrics in terms of the session length distribution and the time cost of user behavior abstraction are displayed in Table~\ref{RQ1 session length static metrics}. \textit{SW (HAC/Markov)} performs better than other simulated workloads in $Mean$ ($18.74\% \rightarrow 18.77\%$, $+0.03\%$) and ${CI}_{0.95}$ ($18.56\% \rightarrow 18.41\%$, $-0.15\%$ in the left boundary and $18.91\% \rightarrow 19.12\%$, $+0.21\%$ in the right boundary), and similar to other simulated workloads in $Min$, $Q1$, $Q2$, $Q3$. However, \textit{SW (HAC/Markov)} has a low value of 4 in $Min$ and a very high value of 155 in $Max$ compared with the original workload. This can be explained by the fact that a few user behavior sequences with low probabilities containing multiple loops can be sampled from the user behavior abstraction as shown in Figure \ref{The abstraction of user behaviors}. According to the values in $Q1$, $Q2$, $Q3$ and ${CI}_{0.95}$ of \textit{SW (HAC/Markov)}, the overall differences between session length distributions are not excessive in relation to the extreme values of $Min$ and $Max$. Besides, each metric in \textit{WESSBAS} almost has a lower value than that in other simulated workloads. This can be explained by the fact that temporal invariants in \textit{WESSBAS} have stricter limits than Synoptic and longer user behavior sequences are more likely to violate. For the think time of the original distribution and the estimated distribution, the area of intersection is around 73.25\% to 73.71\% of $IoU$, which means distributions are highly overlapping. Static metrics in terms of the behavior distribution are displayed in Table~\ref{RQ1 behavior length static metrics}. \textit{SW (HAC/Markov)} has the lowest average difference (0.1267\%) among all relative invocation frequencies. Besides, the total number of distinct session types of \textit{SW (HAC/Markov)} has a minimal increase by 0.15\% ($131415 \rightarrow 131607$) compared with other workloads. Besides, \textit{WESSBAS} still has a low value of 116179 in $N_{ds}$, which also shows fewer long sessions are sampled \textit{WESSBAS} due to the stricter limits. For the time cost of user behavior abstraction, \textit{SW (Monolithic)} has the minimal time cost (12.06s) without clustering and \textit{SW (HAC/Markov)} as well as \textit{WESSBAS} have similar user behavior abstraction time cost (26.99s and 24.34s). Compared with the total time of the whole process (around 9 hours), the time cost of user behavior abstraction is small and acceptable.

\begin{table*}[!t]
\caption{Dynamic metrics between the original workload and the simulated workload in $\mathcal{A}$.}
\label{A dynamic metrics}
\footnotesize
\begin{center}
\begin{tabular}{cccccc}
\hline
Method                           & $ESBD(weak)$ & $score_w$ & $ESBD(strong)$ & $score_s$ & $score_c$ \\ 
\hline
\textit{SW (Monolithic)}         & 0.2297       & 0.8132    & 0.0481         & 0.9541    & 0.9400    \\
\textit{SW (HAC/Markov)}         & 0.2376       & 0.8080    & 0.0436         & 0.9582    & 0.9431    \\
\textit{SW (HAC/SGT)}            & 0.2137       & 0.8239    & 0.0563         & 0.9467    & 0.9345    \\
\textit{WESSBAS}                 & 0.2300       & 0.8130    & 0.1237         & 0.8899    & 0.8822    \\
\hline
\end{tabular}
\end{center}
\vspace{-4mm}
\end{table*}

Table~\ref{A dynamic metrics} shows dynamic metrics between the original workload and the simulated workload in the dataset $\mathcal{A}$. \textit{SW (HAC/Markov)} performs better than other simulated workloads in $ESBD(strong)$ (0.0436), $score_s$ (0.9582), and $score_c$ (0.9431), which also shows \textit{SW (HAC/Markov)} is quite similar with \textit{OW} shows . In metrics weakly correlated with business, \textit{SW (HAC/Markov)} performs worse than other simulated workloads possibly because of the impact on other system processes or applications.

In conclusion, \textit{SW (HAC/Markov)} outperforms other simulated workloads in both static and dynamic metrics, and the similarity between \textit{SW (HAC/Markov)} and \textit{OW} shows the simulated workload generated by LWS can match the original workload accurately.

\begin{figure*}[!t]
\centering
\subfloat[\textit{Fitting} result.]{\includegraphics[width=0.48\textwidth]{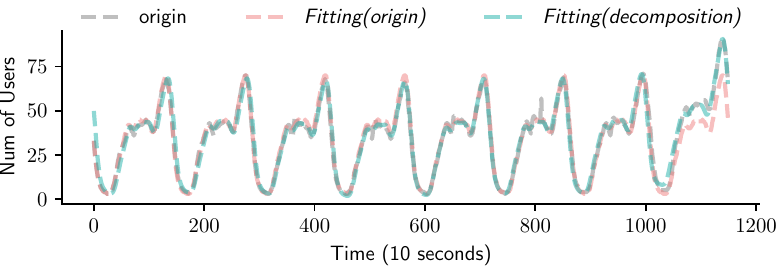}\label{User behavior fitting result}} 
\hfil
\subfloat[\textit{Generation} result.]{\includegraphics[width=0.48\textwidth]{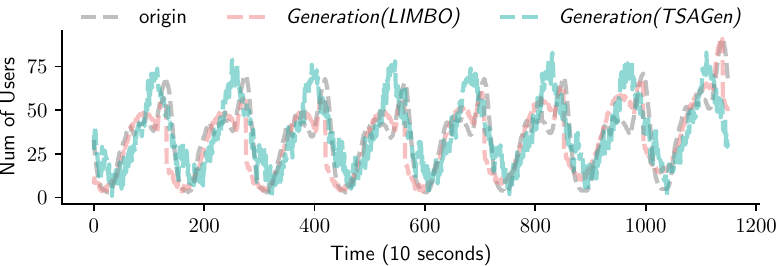}\label{User behavior generation result}} 
\captionsetup{singlelinecheck = false, justification=raggedright, font=footnotesize, labelsep=space}
\caption{User behavior \textit{Fitting} and \textit{Generation} result.}
\vspace{-4mm}
\end{figure*}

\subsection{RQ2: Effectiveness of Intervention}

To evaluate the effectiveness of the intervention, we extract the real-world workload intensity characteristics of the dataset $\mathcal{B}$ and use the fitting method and the generation method to provide interventions. Although both the fitting method and the generation method can provide interventions, their different aims lead to different evaluation criteria. The fitting method learns the characteristics of the original intensity. Therefore, both the similarity between the original intensity and the modeled intensity and the similarity between the original workload and the simulated workload need to be measured. Different from the fitting method, the generation method customizes the intensity from the scratch. Therefore, the similarity between the original intensity and the modeled intensity is not important. It is more important that the generation process is simple and the simulated workload is similar to the real-world workload. Therefore, we use simple parameters to generate the workload similar to the real-world original workload for the evaluation of the generation method. For both the fitting method and the generation method, the original workload is approximated by the workload generated by \textit{Reproduction} as described in Section \ref{dataset} which is used for comparison. The whole process starting from monitoring the application until the simulated workload is acquired is around 9 hours, including the original workload (about 3 hours and 20 minutes), collection and transformation of logs and monitoring metrics (about 1 hour with another 20 minutes waiting for stability in monitoring tools), LWS extraction (about 40 minutes), and the simulated workload (about 3 hours and 20 minutes). In the process of LWS extraction, we have invited 10 non-experts with fundamentals of higher mathematics to fit the original intensity and generate intensity intuitively similar to the original intensity. All of them can finish each task in 30 minutes after reading the 5-minute README file.

\subsubsection{Fitting}

\begin{table*}[!t]
\caption{The overall mathematical expressions of fitting the original intensity in the dataset $\mathcal{B}$.} 
\label{B fitting expressions}
\footnotesize
\begin{center}
\begin{tabular}{c|c|c|c|c|c}
\hline
Method & Component & Expression & SSE & R-square & RMSE \\ 
\hline
\textit{Fitting} & \multirow{2}{*}{-} & \multirow{2}{*}{Fourier} & \multirow{2}{*}{$2.112\times 10^4$} & \multirow{2}{*}{$0.9537$} & \multirow{2}{*}{$4.319$} \\
\textit{(Origin)}& &   &  &  & \\
\hline
& \multirow{1}{*}{$trend$} & \multirow{1}{*}{Polynomial} & \multirow{1}{*}{$79.53$} & \multirow{1}{*}{$0.9922$} & \multirow{1}{*}{$0.264$} \\
\cline{2-6}
\textit{Fitting} & \multirow{1}{*}{$season$} & \multirow{1}{*}{Fourier} & \multirow{1}{*}{$296.8$} & \multirow{1}{*}{$0.9993$} & \multirow{1}{*}{$0.5121$} \\
\cline{2-6}
\textit{(Decomposition)} & \multirow{1}{*}{$noise$} & \multirow{1}{*}{Gaussian} & \multirow{1}{*}{$1.403\times 10^4$} & \multirow{1}{*}{$0.2838$} & \multirow{1}{*}{$3.498$} \\
\cline{2-6}
& \multirow{1}{*}{$merge$} & \multirow{1}{*}{$f(x) = trend+season+noise$} & \multirow{1}{*}{$1.405\times 10^4$} & \multirow{1}{*}{$0.9692$} & \multirow{1}{*}{$3.541$} \\
\hline
\end{tabular}
\end{center}
\vspace{-4mm}
\end{table*}

\begin{table*}[!t]
\caption{Dynamic metrics between the original workload (approximated by \textit{Reproduction}) and the simulated workload in the dataset $\mathcal{B}$.}
\label{B dynamic metrics}
\footnotesize
\begin{center}
\begin{tabular}{cccccc}
\hline
Method                           & $ESBD(weak)$ & $score_w$ & $ESBD(strong)$ & $score_s$ & $score_c$ \\ 
\hline
\textit{Fitting(Origin)}         & 0.1435       & 0.8745    & 0.0123         & 0.9878    & 0.9764    \\
\textit{Fitting(Decomposition)}  & 0.1113       & 0.8998    & 0.0106         & 0.9894    & 0.9805    \\
\textit{Generation(LIMBO)}       & 0.0989       & 0.9099    & 0.0161         & 0.9840    & 0.9766    \\
\textit{Generation(TSAGen)}      & 0.1400       & 0.8771    & 0.0385         & 0.9629    & 0.9543    \\
\hline
\end{tabular}
\end{center}
\vspace{-4mm}
\end{table*}

We use the tool as shown in Figure~\ref{The developed tool for the fitting method} to fit the original intensity. To identify the impact of the decomposition step, both fitting the original intensity directly (denoted by \textit{Fitting(Origin)}) and fitting each part decomposed by RobustSTL (denoted by \textit{Fitting(Decomposition)}) are adopted. Table \ref{B fitting expressions} shows the overall mathematical expressions and evaluation metrics including the sum of squares error (denoted by SSE), R-squared value (denoted by R-square), and the root mean square error (denoted by RMSE) of fitting the original intensity. Each mathematical expression is the most suitable one automatically chosen based on RMSE. For detailed expressions, please refer to supplemental materials in Section \ref{Data availability}. Intensities generated by both \textit{Fitting(Origin)} and \textit{Fitting(Decomposition)} have high R-square (0.9537 and 0.9692), which means that both \textit{Fitting(Origin)} and \textit{Fitting(Decomposition)} can fit the original intensity well. Figure~\ref{User behavior fitting result} shows the original intensity and fitting results. Intensities generated by both \textit{Fitting(Origin)} and \textit{Fitting(Decomposition)} are similar to the original intensity intuitively. The dynamic metrics between the original workload (approximated by \textit{Reproduction}) and the simulated workload whose intensity is generated by \textit{Fitting} are shown in Table~\ref{B dynamic metrics}. Generally, simulated workloads generated by \textit{Fitting(Origin)} and \textit{Fitting(Decomposition)} have low $ESBD(weak)$ (0.1435 \& 0.1113) and $ESBD(strong)$ (0.0123 \& 0.0106) as well as high $score_w$ (0.8745 \& 0.8998), $score_s$ (0.9878 \& 0.9894) and $score_c$ (0.9764 \& 0.9805), which shows the similarity between the original workload and the simulated workload. 

In conclusion, the original intensity and the modeled intensity by \textit{Fitting} are similar, and the corresponding workloads are similar in dynamic metrics as well, which shows that the intervention by \textit{Fitting} is effective. The detailed differences between \textit{Fitting(Origin)} and \textit{Fitting(Decomposition)} will be discussed in Section \ref{RQ3}.

\subsubsection{Generation}

Based on the observation of the characteristics of the original intensity (such as the number of cycles is 8 and each cycle length is 144), we choose simple parameters and interpolation functions such as the linear function and the quadratic function to make the generation process simple and generate the workload intensity as similar as possible to the original intensity. For detailed parameters, please refer to supplemental materials in Section \ref{Data availability}.

It is worth mentioning that the length of the intensity generated by \textit{Generation} is usually different from the original intensity. Therefore, we keep the length of the generated intensity longer than that of the original intensity and select the part whose length is equal to that of the original intensity. Figure~\ref{User behavior generation result} shows the original intensity and generation results. The shapes of the original intensity, the intensity generated by LIMBO, and the intensity generated by TSAGen have intuitive overall similarities. However, compared with fitting results as shown in Figure~\ref{User behavior fitting result}, the intensities generated by \textit{Generation} have more deviations from the original intensity. The dynamic metrics between the original workload (approximated by \textit{Reproduction}) and the simulated workload whose intensity is generated by \textit{Generation} are shown in Table~\ref{B dynamic metrics}. Generally, the original workload and simulated workloads generated by \textit{Generation} still have high-level overall similarities (0.9766 \& 0.9543 in $score_c$). However, there are certain differences 
reflected in $ESBD(strong)$ and $score_s$. We will discuss these differences in Section \ref{RQ3}.

In conclusion, the original workload and the simulated workload whose intensity is generated by \textit{Generation} have high-level overall similarities, which shows that the intervention by \textit{Generation} is effective.

\subsection{RQ3: Impact of Intensity Modeling Methods}\label{RQ3}

We compare different intensity modeling methods by intensity shapes and dynamic metrics of simulated workloads. As shown in Table~\ref{B fitting expressions}, the polynomial model and the Fourier series model fit $trend$ and $season$ in the decomposition pretty well respectively (79.53 and 296.8 in SSE, 0.9922 and 0.9993 in R-square as well as 0.264 and 0.5121 in RMSE). However, the exponential model has poor performance in fitting $noise$ in the decomposition ($1.403\times 10^4$ in SSE, 0.2838 in R-square as well as 3.498 in RMSE), which is the main source of the deviation in \textit{Fitting(Decomposition)}. \textit{Fitting(Origin)} performs worse than \textit{Fitting(Decomposition)} in SSE, R-square and RMSE. The main reason is the strong seasonality of the Fourier series model for \textit{Fitting(Origin)}. As shown in Figure~\ref{User behavior fitting result}, the original intensity has a noticeable increase near the last peak, which cannot be well modeled within the limitation of the Fourier series model. \textit{Fitting(Decomposition)} can model such intensity well due to the flexibility of decomposition. The dynamic metrics of simulated workloads as shown in Table~\ref{B dynamic metrics} also reflect the performance differences. The simulated workload with the \textit{Fitting(Origin)} intensity performs worse than that with the \textit{Fitting(Decomposition)} intensity in all dynamic metrics, which also proves that \textit{Fitting(Decomposition)} is more effective than \textit{Fitting(Origin)}.

For the intensity generated by \textit{Generation}, there is a significant increase in deviations as shown in Figure~\ref{User behavior generation result}. The intensity generated by the TSAGen-based generation method (denoted by \textit{Generation(TSAGen)}) is more irregular, while the intensity generated by the LIMBO-based generation method (denoted by \textit{Generation(LIMBO)}) is smoother. The main reason is that the RMDF process in \textit{Generation(TSAGen)} brings more randomness which is closer to reality and \textit{Generation(LIMBO)} uses interpolation functions with good statistics which are more intuitive and controllable. The simulated workload with the \textit{Generation(TSAGen)} intensity performs worse than that with the \textit{Generation(LIMBO)} intensity in all dynamic metrics as shown in Table~\ref{B dynamic metrics}, which also shows that \textit{Generation(TSAGen)} is more random and uncontrollable compared with \textit{Generation(LIMBO)}.

Although \textit{Fitting} and \textit{Generation} have different aims, both of them can help to generate the simulated workload similar to the original workload. In $ESBD(strong)$, $score_s$ and $score_c$, \textit{Fitting(Decomposition)} still performs better, while \textit{Generation(LIMBO)} even performs better than \textit{Fitting(Origin)}. It is worth mentioning that \textit{Generation} performs better than \textit{Fitting} in $ESBD(weak)$ and $score_w$. The main reason is that there are complex cumulative effects of the simulated workload, other applications, and system events on $ESBD(weak)$ and $score_w$.

In conclusion, \textit{Fitting(Decomposition)} can fit the intensity better than \textit{Fitting(Origin)}, \textit{Generation(LIMBO)} can generate controllable and smooth intensity and \textit{Generation(TSAGen)} can generate irregular intensity with more randomness close to reality. Both \textit{Fitting} and \textit{Generation} can help to generate the effective simulated workload. Besides, we suggest that the original workload should contain typical trends, seasonality, and sudden changes due to their importance in workload intensity modeling. In our practice, the size of logs and metrics in a few hours is at the gigabyte level and suitable to be processed, while more data needs complex tiling and compression technologies. We suggest mapping the original workload into 8 hours or less by undersampling.

\section{Threats to Validity}\label{sec:threats to validity}

The threat to internal validity mainly lies in the implementation of LWS. To reduce this threat, we chose integration over rewriting related tools whenever possible, and carefully checked if the same output can be guaranteed with the same input between the original tool and the rewriting component such as TSAGen. Moreover, we carefully checked and tested the code to ensure that the expected output can be obtained for test cases.

The threat to external validity mainly lies in the benchmark system and the user behavior dataset. We chose an open-source cloud-native microservices application widely used in AIOps tasks and used the load testing script that has been applied in an influential AIOps competition as well as a public dataset containing real user behaviors to produce original workloads. We claim we can use LWS for workload simulation in other session-based systems.

The threat to construct validity mainly lies in the evaluation metrics. We chose two categories of metrics from generation and execution perspectives. Due to the large data size, numerous types, and phase shifts in monitoring metrics, we adopted a KPI-based quality evaluation of workload simulation method proposed in one of our submitted paper, giving a comprehensive evaluation result.

\section{Conclusion and Future Work}\label{sec:conclusion}

AIOps plays a critical role in the operation and management of cloud-native systems and microservice-based applications. However, the lack of high-quality datasets with diverse scenarios constrains the development and evaluation of AIOps research. Workload simulation that produces realistic workloads is an important task for continually generating such AIOps datasets. In this paper, we formulate the task of workload simulation and propose a novel framework, LWS, for log-based workload simulation in the session-based system. LWS consists of four components, namely log collection and transformation, user behavior abstraction, workload intensity modeling, and simulated workload generation. Hierarchical clustering based on Markov probabilistic transition matrix is applied to identify user groups and the relational model based on temporal invariants for each group is intended for user behavior abstraction. Three workload intensity modeling methods, namely \textit{Reproduction}, \textit{Fitting}, and \textit{Generation}, are introduced in workload intensity modeling for better intervention. Experiment results on a typical cloud-native microservices application with both the well-designed workload and the real-world workload show that the simulated workload generated by LWS is effective and intervenable.

In our future work, we will investigate the automatic selection of temporal invariants in the relational model in the user behavior abstraction. Moreover, we will explore workload simulation with failures induced by external injection through Chaos engineering or internal orchestration through well-designed business workflows that include faults, which can enrich the failure scenarios and further extend our motivation of generating high-quality AIOps datasets.

\section{Data Availability}\label{Data availability}

The supplemental materials can be accessed through \url{https://github.com/baiyanquan/LWS}. The data of this work can be accessed through \url{https://figshare.com/s/87c50b19706242dc0f1e}. The source code of this work is under limitations of non-disclosure agreements in the Beijing BizSeer Technology Company. Please contact us via the email contact@bizseer.com or the telephone number 010-82362970 if needed.

\section{Acknowledgement}

This work was supported by National Key R\&D Program of
China (Grant No. 2020YFB2103300). We also gratefully acknowledge the support of the Beijing BizSeer Technology Company.


\begin{thebibliography}{49}
\expandafter\ifx\csname natexlab\endcsname\relax\def\natexlab#1{#1}\fi
\providecommand{\url}[1]{\texttt{#1}}
\providecommand{\href}[2]{#2}
\providecommand{\path}[1]{#1}
\providecommand{\DOIprefix}{doi:}
\providecommand{\ArXivprefix}{arXiv:}
\providecommand{\URLprefix}{URL: }
\providecommand{\Pubmedprefix}{pmid:}
\providecommand{\doi}[1]{\href{http://dx.doi.org/#1}{\path{#1}}}
\providecommand{\Pubmed}[1]{\href{pmid:#1}{\path{#1}}}
\providecommand{\bibinfo}[2]{#2}
\ifx\xfnm\relax \def\xfnm[#1]{\unskip,\space#1}\fi
\bibitem[{Abbors et~al.(2012)Abbors, Ahmad, Truscan and
  Porres}]{abbors2012mbpet}
\bibinfo{author}{Abbors, F.}, \bibinfo{author}{Ahmad, T.},
  \bibinfo{author}{Truscan, D.}, \bibinfo{author}{Porres, I.},
  \bibinfo{year}{2012}.
\newblock \bibinfo{title}{{MBPeT: A model-based performance testing tool}}, in:
  \bibinfo{booktitle}{VALID 2012 - 4th International Conference on Advances in
  System Testing and Validation Lifecycle}, pp. \bibinfo{pages}{1--8}.
\bibitem[{Barnert and Krcmar(2021)}]{barnert2021simulation}
\bibinfo{author}{Barnert, M.}, \bibinfo{author}{Krcmar, H.},
  \bibinfo{year}{2021}.
\newblock \bibinfo{title}{{Simulation of in-memory database workload: Markov
  chains versus relative invocation frequency and equal probability - A
  trade-off between accuracy and time}}, in: \bibinfo{booktitle}{Proceedings of
  the ACM/SPEC International Conference on Performance Engineering}, p.
  \bibinfo{pages}{73–80}.
\newblock \DOIprefix\doi{10.1145/3427921.3450237}.
\bibitem[{Biermann and Feldman(1972)}]{biermann1972synthesis}
\bibinfo{author}{Biermann, A.W.}, \bibinfo{author}{Feldman, J.A.},
  \bibinfo{year}{1972}.
\newblock \bibinfo{title}{{On the synthesis of finite-state machines from
  samples of their behavior}}.
\newblock \bibinfo{journal}{IEEE Transactions on Computers}
  \bibinfo{volume}{C-21}, \bibinfo{pages}{592--597}.
\newblock \DOIprefix\doi{10.1109/TC.1972.5009015}.
\bibitem[{Calzarossa et~al.(2016)Calzarossa, Massari and
  Tessera}]{calzarossa2016workload}
\bibinfo{author}{Calzarossa, M.C.}, \bibinfo{author}{Massari, L.},
  \bibinfo{author}{Tessera, D.}, \bibinfo{year}{2016}.
\newblock \bibinfo{title}{{Workload characterization: A survey revisited}}.
\newblock \bibinfo{journal}{ACM Computing Surveys} \bibinfo{volume}{48},
  \bibinfo{pages}{1–43}.
\newblock \DOIprefix\doi{10.1145/2856127}.
\bibitem[{Curiel and Pont(2018)}]{curiel2018workload}
\bibinfo{author}{Curiel, M.}, \bibinfo{author}{Pont, A.}, \bibinfo{year}{2018}.
\newblock \bibinfo{title}{{Workload generators for web-based systems:
  Characteristics, current status, and challenges}}.
\newblock \bibinfo{journal}{IEEE Communications Surveys \& Tutorials}
  \bibinfo{volume}{20}, \bibinfo{pages}{1526--1546}.
\newblock \DOIprefix\doi{10.1109/COMST.2018.2798641}.
\bibitem[{Dennis and Mor\'{e}(1977)}]{dennis1977quasi}
\bibinfo{author}{Dennis, Jr., J.E.}, \bibinfo{author}{Mor\'{e}, J.J.},
  \bibinfo{year}{1977}.
\newblock \bibinfo{title}{{Quasi-Newton methods, motivation and theory}}.
\newblock \bibinfo{journal}{SIAM Review} \bibinfo{volume}{19},
  \bibinfo{pages}{46--89}.
\newblock \DOIprefix\doi{10.1137/1019005}.
\bibitem[{Draheim et~al.(2006)Draheim, Grundy, Hosking, Lutteroth and
  Weber}]{Draheim2006RealisticLT}
\bibinfo{author}{Draheim, D.}, \bibinfo{author}{Grundy, J.},
  \bibinfo{author}{Hosking, J.}, \bibinfo{author}{Lutteroth, C.},
  \bibinfo{author}{Weber, G.}, \bibinfo{year}{2006}.
\newblock \bibinfo{title}{{Realistic load testing of web applications}}, in:
  \bibinfo{booktitle}{Conference on Software Maintenance and Reengineering
  (CSMR'06)}, pp. \bibinfo{pages}{57--67}.
\newblock \DOIprefix\doi{10.1109/CSMR.2006.43}.
\bibitem[{Erradi et~al.(2019)Erradi, Iqbal, Mahmood and
  Bouguettaya}]{erradi2019web}
\bibinfo{author}{Erradi, A.}, \bibinfo{author}{Iqbal, W.},
  \bibinfo{author}{Mahmood, A.}, \bibinfo{author}{Bouguettaya, A.},
  \bibinfo{year}{2019}.
\newblock \bibinfo{title}{{Web application resource requirements estimation
  based on the workload latent features}}.
\newblock \bibinfo{journal}{IEEE Transactions on Services Computing}
  \bibinfo{volume}{14}, \bibinfo{pages}{1638--1649}.
\newblock \DOIprefix\doi{10.1109/TSC.2019.2918776}.
\bibitem[{Fattah et~al.(2020)Fattah, Bouguettaya and Mistry}]{fattah2020long}
\bibinfo{author}{Fattah, S.M.M.}, \bibinfo{author}{Bouguettaya, A.},
  \bibinfo{author}{Mistry, S.}, \bibinfo{year}{2020}.
\newblock \bibinfo{title}{{Long-term IaaS selection using performance
  discovery}}.
\newblock \bibinfo{journal}{IEEE Transactions on Services Computing}
  \bibinfo{volume}{15}, \bibinfo{pages}{2129--2143}.
\newblock \DOIprefix\doi{10.1109/TSC.2020.3036677}.
\bibitem[{Fei et~al.(2020)Fei, Zhu, Liu, Chen, Bao and Liu}]{fei2020elastic}
\bibinfo{author}{Fei, B.}, \bibinfo{author}{Zhu, X.}, \bibinfo{author}{Liu,
  D.}, \bibinfo{author}{Chen, J.}, \bibinfo{author}{Bao, W.},
  \bibinfo{author}{Liu, L.}, \bibinfo{year}{2020}.
\newblock \bibinfo{title}{{Elastic resource provisioning using data clustering
  in cloud service platform}}.
\newblock \bibinfo{journal}{IEEE Transactions on Services Computing}
  \bibinfo{volume}{15}, \bibinfo{pages}{1578--1591}.
\newblock \DOIprefix\doi{10.1109/TSC.2020.3002755}.
\bibitem[{Feng et~al.(2022)Feng, Ding and Jiang}]{feng2022fast}
\bibinfo{author}{Feng, B.}, \bibinfo{author}{Ding, Z.}, \bibinfo{author}{Jiang,
  C.}, \bibinfo{year}{2022}.
\newblock \bibinfo{title}{{FAST: A forecasting model with adaptive sliding
  window and time locality integration for dynamic cloud workloads}}.
\newblock \bibinfo{journal}{IEEE Transactions on Services Computing}
  \DOIprefix\doi{10.1109/TSC.2022.3156619}.
\bibitem[{Fournier et~al.(1982)Fournier, Fussell and
  Carpenter}]{fournier1982computer}
\bibinfo{author}{Fournier, A.}, \bibinfo{author}{Fussell, D.},
  \bibinfo{author}{Carpenter, L.}, \bibinfo{year}{1982}.
\newblock \bibinfo{title}{{Computer rendering of stochastic models}}.
\newblock \bibinfo{journal}{Communications of the ACM} \bibinfo{volume}{25},
  \bibinfo{pages}{371–384}.
\newblock \DOIprefix\doi{10.1145/358523.358553}.
\bibitem[{Goeva-Popstojanova et~al.(2006)Goeva-Popstojanova, Singh, Mazimdar
  and Li}]{GosevaPopstojanova2006EmpiricalCO}
\bibinfo{author}{Goeva-Popstojanova, K.}, \bibinfo{author}{Singh, A.D.},
  \bibinfo{author}{Mazimdar, S.}, \bibinfo{author}{Li, F.},
  \bibinfo{year}{2006}.
\newblock \bibinfo{title}{{Empirical characterization of session-based workload
  and reliability for Web servers}}.
\newblock \bibinfo{journal}{Empirical Software Engineering}
  \bibinfo{volume}{11}, \bibinfo{pages}{71--117}.
\newblock \DOIprefix\doi{10.1007/s10664-006-5966-7}.
\bibitem[{Goldstein et~al.(2017)Goldstein, Raz and
  Segall}]{goldstein2017experience}
\bibinfo{author}{Goldstein, M.}, \bibinfo{author}{Raz, D.},
  \bibinfo{author}{Segall, I.}, \bibinfo{year}{2017}.
\newblock \bibinfo{title}{{Experience report: Log-based behavioral
  differencing}}, in: \bibinfo{booktitle}{2017 IEEE 28th International
  Symposium on Software Reliability Engineering (ISSRE)}, pp.
  \bibinfo{pages}{282--293}.
\newblock \DOIprefix\doi{10.1109/ISSRE.2017.14}.
\bibitem[{Gu et~al.(2020)Gu, Ding, Wang and Yin}]{gu2020hierarchical}
\bibinfo{author}{Gu, Y.}, \bibinfo{author}{Ding, Z.}, \bibinfo{author}{Wang,
  S.}, \bibinfo{author}{Yin, D.}, \bibinfo{year}{2020}.
\newblock \bibinfo{title}{{Hierarchical User Profiling for E-Commerce
  Recommender Systems}}, in: \bibinfo{booktitle}{Proceedings of the 13th
  International Conference on Web Search and Data Mining}, p.
  \bibinfo{pages}{223–231}.
\newblock \DOIprefix\doi{10.1145/3336191.3371827}.
\bibitem[{Herbst et~al.(2014)Herbst, Huber, Kounev and Amrehn}]{herbst2014self}
\bibinfo{author}{Herbst, N.R.}, \bibinfo{author}{Huber, N.},
  \bibinfo{author}{Kounev, S.}, \bibinfo{author}{Amrehn, E.},
  \bibinfo{year}{2014}.
\newblock \bibinfo{title}{{Self-adaptive workload classification and
  forecasting for proactive resource provisioning}}.
\newblock \bibinfo{journal}{Concurrency and Computation: Practice and
  Experience} \bibinfo{volume}{26}, \bibinfo{pages}{2053--2078}.
\newblock \DOIprefix\doi{10.1002/cpe.3224}.
\bibitem[{Kang et~al.(2010)Kang, Shin and Shin}]{kang2010detecting}
\bibinfo{author}{Kang, W.}, \bibinfo{author}{Shin, D.}, \bibinfo{author}{Shin,
  D.}, \bibinfo{year}{2010}.
\newblock \bibinfo{title}{{Detecting and predicting of abnormal behavior using
  hierarchical Markov model in smart home network}}, in:
  \bibinfo{booktitle}{2010 IEEE 17Th International Conference on Industrial
  Engineering and Engineering Management}, pp. \bibinfo{pages}{410--414}.
\newblock \DOIprefix\doi{10.1109/ICIEEM.2010.5646583}.
\bibitem[{v.~Kistowski et~al.(2014)v.~Kistowski, Herbst and
  Kounev}]{v2014modeling}
\bibinfo{author}{v.~Kistowski, J.}, \bibinfo{author}{Herbst, N.R.},
  \bibinfo{author}{Kounev, S.}, \bibinfo{year}{2014}.
\newblock \bibinfo{title}{{Modeling variations in load intensity over time}},
  in: \bibinfo{booktitle}{Proceedings of the Third International Workshop on
  Large Scale Testing}, pp. \bibinfo{pages}{1--4}.
\newblock \DOIprefix\doi{10.1145/2577036.2577037}.
\bibitem[{Kistowski et~al.(2017)Kistowski, Herbst, Kounev, Groenda, Stier and
  Lehrig}]{kistowski2017modeling}
\bibinfo{author}{Kistowski, J.V.}, \bibinfo{author}{Herbst, N.},
  \bibinfo{author}{Kounev, S.}, \bibinfo{author}{Groenda, H.},
  \bibinfo{author}{Stier, C.}, \bibinfo{author}{Lehrig, S.},
  \bibinfo{year}{2017}.
\newblock \bibinfo{title}{{Modeling and extracting load intensity profiles}}.
\newblock \bibinfo{journal}{ACM Transactions on Autonomous and Adaptive Systems
  (TAAS)} \bibinfo{volume}{11}, \bibinfo{pages}{1--28}.
\newblock \DOIprefix\doi{10.1145/3019596}.
\bibitem[{Kratzke and Quint(2017)}]{kratzke2017understanding}
\bibinfo{author}{Kratzke, N.}, \bibinfo{author}{Quint, P.C.},
  \bibinfo{year}{2017}.
\newblock \bibinfo{title}{{Understanding cloud-native applications after 10
  years of cloud computing - A systematic mapping study}}.
\newblock \bibinfo{journal}{Journal of Systems and Software}
  \bibinfo{volume}{126}, \bibinfo{pages}{1--16}.
\newblock \DOIprefix\doi{10.1016/j.jss.2017.01.001}.
\bibitem[{Lee et~al.(2011)Lee, Lo and Fu}]{lee2011novel}
\bibinfo{author}{Lee, C.H.}, \bibinfo{author}{Lo, Y.l.}, \bibinfo{author}{Fu,
  Y.H.}, \bibinfo{year}{2011}.
\newblock \bibinfo{title}{{A novel prediction model based on hierarchical
  characteristic of web site}}.
\newblock \bibinfo{journal}{Expert Systems with Applications}
  \bibinfo{volume}{38}, \bibinfo{pages}{3422--3430}.
\newblock \DOIprefix\doi{10.1016/j.eswa.2010.08.128}.
\bibitem[{Li et~al.(2023)Li, Du and Zhao}]{workloadsimulationevaluation}
\bibinfo{author}{Li, P.}, \bibinfo{author}{Du, Q.}, \bibinfo{author}{Zhao, S.},
  \bibinfo{year}{2023}.
\newblock \bibinfo{title}{{KEWS: a method evaluation of workload simulation
  based on KPIs}}.
\newblock \URLprefix \url{https://arxiv.org/abs/2301.06530},
  \DOIprefix\doi{10.48550/ARXIV.2301.06530}.
\bibitem[{Li and Tian(2003)}]{Li2003TestingTS}
\bibinfo{author}{Li, Z.}, \bibinfo{author}{Tian, J.}, \bibinfo{year}{2003}.
\newblock \bibinfo{title}{{Testing the suitability of Markov chains as web
  usage models}}, in: \bibinfo{booktitle}{Proceedings 27th Annual International
  Computer Software and Applications Conference. COMPAC 2003}, pp.
  \bibinfo{pages}{356--361}.
\newblock \DOIprefix\doi{10.1109/CMPSAC.2003.1245365}.
\bibitem[{Li et~al.(2022)Li, Zhao, Zhang, Sun, Chen, Wen, Ma and
  Pei}]{li2022constructing}
\bibinfo{author}{Li, Z.}, \bibinfo{author}{Zhao, N.}, \bibinfo{author}{Zhang,
  S.}, \bibinfo{author}{Sun, Y.}, \bibinfo{author}{Chen, P.},
  \bibinfo{author}{Wen, X.}, \bibinfo{author}{Ma, M.}, \bibinfo{author}{Pei,
  D.}, \bibinfo{year}{2022}.
\newblock \bibinfo{title}{{Constructing large-scale real-world benchmark
  datasets for AIOps}}.
\newblock \DOIprefix\doi{10.48550/arXiv.2208.03938}.
\bibitem[{Lutteroth and Weber(2008)}]{lutteroth2008modeling}
\bibinfo{author}{Lutteroth, C.}, \bibinfo{author}{Weber, G.},
  \bibinfo{year}{2008}.
\newblock \bibinfo{title}{{Modeling a realistic workload for performance
  testing}}, in: \bibinfo{booktitle}{2008 12th International IEEE Enterprise
  Distributed Object Computing Conference}, pp. \bibinfo{pages}{149--158}.
\newblock \DOIprefix\doi{10.1109/EDOC.2008.40}.
\bibitem[{Menasc\'{e} et~al.(1999)Menasc\'{e}, Almeida, Fonseca and
  Mendes}]{Menasc1999AMF}
\bibinfo{author}{Menasc\'{e}, D.A.}, \bibinfo{author}{Almeida, V.A.F.},
  \bibinfo{author}{Fonseca, R.}, \bibinfo{author}{Mendes, M.A.},
  \bibinfo{year}{1999}.
\newblock \bibinfo{title}{{A methodology for workload characterization of
  e-commerce sites}}, in: \bibinfo{booktitle}{Proceedings of the 1st ACM
  Conference on Electronic Commerce}, p. \bibinfo{pages}{119–128}.
\newblock \DOIprefix\doi{10.1145/336992.337024}.
\bibitem[{Notaro et~al.(2021)Notaro, Cardoso and Gerndt}]{notaro2021survey}
\bibinfo{author}{Notaro, P.}, \bibinfo{author}{Cardoso, J.},
  \bibinfo{author}{Gerndt, M.}, \bibinfo{year}{2021}.
\newblock \bibinfo{title}{{A survey of AIOps methods for failure management}}.
\newblock \bibinfo{journal}{ACM Transactions on Intelligent Systems and
  Technology (TIST)} \bibinfo{volume}{12}, \bibinfo{pages}{1--45}.
\newblock \DOIprefix\doi{10.1145/3483424}.
\bibitem[{{\"o}gele et~al.(2018){\"o}gele, van Hoorn, Schulz, Hasselbring and
  Krcmar}]{vogele2018wessbas}
\bibinfo{author}{V{\"o}gele, C.}, \bibinfo{author}{van Hoorn, A.},
  \bibinfo{author}{Schulz, E.}, \bibinfo{author}{Hasselbring, W.},
  \bibinfo{author}{Krcmar, H.}, \bibinfo{year}{2018}.
\newblock \bibinfo{title}{{WESSBAS: Extraction of probabilistic workload
  specifications for load testing and performance prediction---a model-driven
  approach for session-based application systems}}.
\newblock \bibinfo{journal}{Software {\&} Systems Modeling}
  \bibinfo{volume}{17}, \bibinfo{pages}{443--477}.
\newblock \DOIprefix\doi{10.1007/s10270-016-0566-5}.
\bibitem[{Ohmann et~al.(2014)Ohmann, Herzberg, Fiss, Halbert, Palyart,
  Beschastnikh and Brun}]{ohmann2014behavioral}
\bibinfo{author}{Ohmann, T.}, \bibinfo{author}{Herzberg, M.},
  \bibinfo{author}{Fiss, S.}, \bibinfo{author}{Halbert, A.},
  \bibinfo{author}{Palyart, M.}, \bibinfo{author}{Beschastnikh, I.},
  \bibinfo{author}{Brun, Y.}, \bibinfo{year}{2014}.
\newblock \bibinfo{title}{{Behavioral resource-aware model inference}}, in:
  \bibinfo{booktitle}{Proceedings of the 29th ACM/IEEE International Conference
  on Automated Software Engineering}, p. \bibinfo{pages}{19–30}.
\newblock \DOIprefix\doi{10.1145/2642937.2642988}.
\bibitem[{Paparrizos and Gravano(2015)}]{paparrizos2015k}
\bibinfo{author}{Paparrizos, J.}, \bibinfo{author}{Gravano, L.},
  \bibinfo{year}{2015}.
\newblock \bibinfo{title}{{K-Shape: Efficient and accurate clustering of time
  series}}, in: \bibinfo{booktitle}{Proceedings of the 2015 ACM SIGMOD
  International Conference on Management of Data}, p.
  \bibinfo{pages}{1855–1870}.
\newblock \DOIprefix\doi{10.1145/2949741.2949758}.
\bibitem[{Parrott and Carver(2020)}]{parrott2020lodeston}
\bibinfo{author}{Parrott, C.}, \bibinfo{author}{Carver, D.},
  \bibinfo{year}{2020}.
\newblock \bibinfo{title}{{Lodestone: A streaming approach to behavior modeling
  and load testing}}, in: \bibinfo{booktitle}{2020 3rd International Conference
  on Data Intelligence and Security (ICDIS)}, pp. \bibinfo{pages}{109--116}.
\newblock \DOIprefix\doi{10.1109/ICDIS50059.2020.00021}.
\bibitem[{Ranjan et~al.(2022)Ranjan, Ebrahimi and
  Paynabar}]{ranjan2022sequence}
\bibinfo{author}{Ranjan, C.}, \bibinfo{author}{Ebrahimi, S.},
  \bibinfo{author}{Paynabar, K.}, \bibinfo{year}{2022}.
\newblock \bibinfo{title}{{Sequence graph transform (SGT): a feature embedding
  function for sequence data mining}}.
\newblock \bibinfo{journal}{Data Mining and Knowledge Discovery}
  \bibinfo{volume}{36}, \bibinfo{pages}{668--708}.
\newblock \DOIprefix\doi{10.1007/s10618-021-00813-0}.
\bibitem[{Reiss et~al.(2012)Reiss, Tumanov, Ganger, Katz and
  Kozuch}]{reiss2012heterogeneity}
\bibinfo{author}{Reiss, C.}, \bibinfo{author}{Tumanov, A.},
  \bibinfo{author}{Ganger, G.R.}, \bibinfo{author}{Katz, R.H.},
  \bibinfo{author}{Kozuch, M.A.}, \bibinfo{year}{2012}.
\newblock \bibinfo{title}{{Heterogeneity and dynamicity of clouds at scale:
  Google trace analysis}}, in: \bibinfo{booktitle}{Proceedings of the Third ACM
  Symposium on Cloud Computing}, pp. \bibinfo{pages}{1--13}.
\newblock \DOIprefix\doi{10.1145/2391229.2391236}.
\bibitem[{Ruffo et~al.(2004)Ruffo, Schifanella, Sereno and
  Politi}]{Ruffo2004WALTyAU}
\bibinfo{author}{Ruffo, G.}, \bibinfo{author}{Schifanella, R.},
  \bibinfo{author}{Sereno, M.}, \bibinfo{author}{Politi, R.},
  \bibinfo{year}{2004}.
\newblock \bibinfo{title}{{WALTy: a user behavior tailored tool for evaluating
  Web application performance}}, in: \bibinfo{booktitle}{Third IEEE
  International Symposium on Network Computing and Applications, 2004. (NCA
  2004). Proceedings.}, pp. \bibinfo{pages}{77--86}.
\newblock \DOIprefix\doi{10.1109/NCA.2004.1347765}.
\bibitem[{Schneider et~al.(2010)Schneider, Beschastnikh, Chernyak, Ernst and
  Brun}]{schneider2010synoptic}
\bibinfo{author}{Schneider, S.}, \bibinfo{author}{Beschastnikh, I.},
  \bibinfo{author}{Chernyak, S.}, \bibinfo{author}{Ernst, M.D.},
  \bibinfo{author}{Brun, Y.}, \bibinfo{year}{2010}.
\newblock \bibinfo{title}{{Synoptic: Summarizing system logs with refinement}},
  in: \bibinfo{booktitle}{Proceedings of the 2010 Workshop on Managing Systems
  via Log Analysis and Machine Learning Techniques}, p.~\bibinfo{pages}{2}.
\newblock \DOIprefix\doi{10.1145/1928991.1928995}.
\bibitem[{Schulz et~al.(2019)Schulz, Angerstein, Okanović and van
  Hoorn}]{Schulz2019MicroserviceTailoredGO}
\bibinfo{author}{Schulz, H.}, \bibinfo{author}{Angerstein, T.},
  \bibinfo{author}{Okanović, D.}, \bibinfo{author}{van Hoorn, A.},
  \bibinfo{year}{2019}.
\newblock \bibinfo{title}{{Microservice-tailored generation of session-based
  workload models for representative load testing}}, in:
  \bibinfo{booktitle}{2019 IEEE 27th International Symposium on Modeling,
  Analysis, and Simulation of Computer and Telecommunication Systems
  (MASCOTS)}, pp. \bibinfo{pages}{323--335}.
\newblock \DOIprefix\doi{10.1109/MASCOTS.2019.00043}.
\bibitem[{Schulz et~al.(2021)Schulz, Okanovi\'{c}, van Hoorn and
  T\r{u}ma}]{schulz2021context}
\bibinfo{author}{Schulz, H.}, \bibinfo{author}{Okanovi\'{c}, D.},
  \bibinfo{author}{van Hoorn, A.}, \bibinfo{author}{T\r{u}ma, P.},
  \bibinfo{year}{2021}.
\newblock \bibinfo{title}{{Context-tailored workload model generation for
  continuous representative load testing}}, in: \bibinfo{booktitle}{Proceedings
  of the ACM/SPEC International Conference on Performance Engineering}, p.
  \bibinfo{pages}{21–32}.
\newblock \DOIprefix\doi{10.1145/3427921.3450240}.
\bibitem[{Shams et~al.(2006)Shams, Krishnamurthy and Far}]{Shams2006AMA}
\bibinfo{author}{Shams, M.}, \bibinfo{author}{Krishnamurthy, D.},
  \bibinfo{author}{Far, B.}, \bibinfo{year}{2006}.
\newblock \bibinfo{title}{{A model-based approach for testing the performance
  of web applications}}, in: \bibinfo{booktitle}{Proceedings of the 3rd
  International Workshop on Software Quality Assurance}, p.
  \bibinfo{pages}{54–61}.
\newblock \DOIprefix\doi{10.1145/1188895.1188909}.
\bibitem[{Silverman(1998)}]{KDE}
\bibinfo{author}{Silverman, B.W.}, \bibinfo{year}{1998}.
\newblock \bibinfo{title}{{Density estimation for statistics and data
  analysis}}.
\newblock \bibinfo{publisher}{Routledge}.
\newblock \DOIprefix\doi{10.1201/9781315140919}.
\bibitem[{Soldani et~al.(2018)Soldani, Tamburri and Van
  Den~Heuvel}]{soldani2018pains}
\bibinfo{author}{Soldani, J.}, \bibinfo{author}{Tamburri, D.A.},
  \bibinfo{author}{Van Den~Heuvel, W.J.}, \bibinfo{year}{2018}.
\newblock \bibinfo{title}{{The pains and gains of microservices: A Systematic grey literature review}}.
\newblock \bibinfo{journal}{Journal of Systems and Software}
  \bibinfo{volume}{146}, \bibinfo{pages}{215--232}.
\newblock \DOIprefix\doi{10.1016/j.jss.2018.09.082}.
\bibitem[{Taylor and Letham(2018)}]{taylor2018forecasting}
\bibinfo{author}{Taylor, S.J.}, \bibinfo{author}{Letham, B.},
  \bibinfo{year}{2018}.
\newblock \bibinfo{title}{{Forecasting at scale}}.
\newblock \bibinfo{journal}{The American Statistician} \bibinfo{volume}{72},
  \bibinfo{pages}{37--45}.
\newblock \DOIprefix\doi{10.1080/00031305.2017.1380080}.
\bibitem[{Wang et~al.(2021)Wang, Wu, Zhou, Yu and Cai}]{wang2021tsagen}
\bibinfo{author}{Wang, C.}, \bibinfo{author}{Wu, K.}, \bibinfo{author}{Zhou,
  T.}, \bibinfo{author}{Yu, G.}, \bibinfo{author}{Cai, Z.},
  \bibinfo{year}{2021}.
\newblock \bibinfo{title}{{TSAGen: Synthetic time series generation for KPI
  anomaly detection}}.
\newblock \bibinfo{journal}{IEEE Transactions on Network and Service
  Management} \bibinfo{volume}{19}, \bibinfo{pages}{130--145}.
\newblock \DOIprefix\doi{10.1109/TNSM.2021.3098784}.
\bibitem[{Wen et~al.(2019)Wen, Gao, Song, Sun, Xu and
  Zhu}]{DBLP:conf/aaai/WenGS0XZ19}
\bibinfo{author}{Wen, Q.}, \bibinfo{author}{Gao, J.}, \bibinfo{author}{Song,
  X.}, \bibinfo{author}{Sun, L.}, \bibinfo{author}{Xu, H.},
  \bibinfo{author}{Zhu, S.}, \bibinfo{year}{2019}.
\newblock \bibinfo{title}{{RobustSTL: A robust seasonal-trend decomposition
  algorithm for long time series}}, in: \bibinfo{booktitle}{33rd AAAI
  Conference on Artificial Intelligence, AAAI 2019, 31st Innovative
  Applications of Artificial Intelligence Conference, IAAI 2019 and the 9th
  AAAI Symposium on Educational Advances in Artificial Intelligence, EAAI
  2019}, pp. \bibinfo{pages}{5409--5416}.
\newblock \DOIprefix\doi{10.1609/aaai.v33i01.33015409}.
\bibitem[{Xu and Wunsch(2005)}]{xu2005survey}
\bibinfo{author}{Xu, R.}, \bibinfo{author}{Wunsch, D.}, \bibinfo{year}{2005}.
\newblock \bibinfo{title}{{Survey of clustering algorithms}}.
\newblock \bibinfo{journal}{IEEE Transactions on Neural Networks}
  \bibinfo{volume}{16}, \bibinfo{pages}{645--678}.
\newblock \DOIprefix\doi{10.1109/TNN.2005.845141}.
\bibitem[{Zhou et~al.(2014)Zhou, Zhou and Li}]{Zhou2014LTFAM}
\bibinfo{author}{Zhou, J.}, \bibinfo{author}{Zhou, B.}, \bibinfo{author}{Li,
  S.}, \bibinfo{year}{2014}.
\newblock \bibinfo{title}{{LTF: A model-based load testing framework for web
  applications}}, in: \bibinfo{booktitle}{2014 14th International Conference on
  Quality Software}, pp. \bibinfo{pages}{154--163}.
\newblock \DOIprefix\doi{10.1109/QSIC.2014.53}.

\end{thebibliography}
\end{document}